\documentclass[11pt,a4paper]{article}
\usepackage[left=3cm, right=3cm, top=3.5cm]{geometry}
\usepackage{amssymb,amsmath, amsthm, amsfonts, rotating}
\usepackage{mathrsfs,color,bbold,url}
\usepackage{stmaryrd}
\usepackage{tikz}
\usepackage[all]{xypic}
\usepackage{enumerate}

\newtheorem{theorem}{Theorem}[section]
\newtheorem{lemma}[theorem]{Lemma}
\newtheorem{corollary}[theorem]{Corollary}
\newtheorem{proposition}[theorem]{Proposition}

\theoremstyle{definition}
\newtheorem{definition}[theorem]{Definition}

\newtheorem{example}[theorem]{Example}

\usepackage{marvosym}
\usepackage{nicefrac}
\usepackage{circuitikz}
\usepackage{adjustbox}
\usepackage[most]{tcolorbox}

\usepackage{tikz-qtree}
\usepackage{tikz-qtree-compat}

\tikzset{every tree node/.style={align=center,anchor=north}}

\makeatother

\begin{document}

\title{A First Polynomial Non-Clausal Class in   Many-Valued Logic}
\author{Gonzalo E. Imaz\\ [2mm] %
{\small  Artificial Intelligence Research Institute (IIIA) - CSIC, Barcelona, Spain}\\
{\small \tt email: {\{gonzalo\}}@iiia.csic.es}\\[1mm]
}

\date{}
\maketitle

\begin{abstract}

The relevance of  polynomial formula classes to deductive efficiency 
 motivated their search,    and currently, 
  a great  number of such classes is known.
Nonetheless, they have  been exclusively sought  in the  setting of
  {\em clausal form} and propositional logic, which  is of course expressively
 limiting for  real applications. As a consequence,  a first polynomial 
  propositional  class in {\em non-clausal (NC) form} has recently been  proposed.

Along these lines  and towards   making  NC tractability applicable 
beyond propositional logic, firstly, we  {\em   define
   the  $\mathcal{R}$egular many-valued $\mathcal{H}$orn Non-Clausal class,} 
   or $\mathcal{RH}$, obtained   by 
 suitably amalgamating  both regular classes:  Horn 
 and  NC. 
 
 Secondly, we demonstrate that  the relationship between  (1)  
 $\mathcal{RH}$ and  the regular Horn class   is that syntactically $\mathcal{RH}$  subsumes the Horn class but  
  that  both  classes   are    equivalent semantically; and between (2) 
   $\mathcal{RH}$ and the regular non-clausal class
     is that  $\mathcal{RH}$ contains all NC formulas whose clausal form is   Horn.
 
 Thirdly,  we define    
  {\em  Regular Non-Clausal   Unit-Resolution,} 
  or $\mathit{RUR_{NC}}$, and prove
  both that it is   complete for $\mathcal{RH}$ and  that
  checks its satisfiability in  polynomial time. The latter fact shows
  that  our intended goal is reached since $\mathcal{RH}$ is  
    many-valued, non-clausal and tractable.
 
 As $\mathcal{RH}$ and $\mathit{RUR_{NC}}$ 
 are, both, basic in the DPLL scheme, the most efficient  in propositional logic, and  can be extended
 to some other   non-classical logics, we argue that they 
 pave the way for efficient non-clausal DPLL-based 
 approximate  reasoning.

\vspace{.15cm}   
\noindent  {\bf Field:}   {\em Tractable Approximate Automated Reasoning.}
 
\vspace{.05cm} 
\end{abstract}



\newenvironment{niceproof}{\trivlist\item[\hskip
\labelsep{\it Proof.\/}]\ignorespaces}{\hfill$\blacksquare$\endtrivlist}

\newenvironment{proofsketch}{\trivlist\item[\hskip
\labelsep{\it Proof Sketch.\/}]\ignorespaces}{\hfill$\blacksquare$\endtrivlist}

\newenvironment{niceproofsketch}{\trivlist\item[\hskip
       \labelsep{\bf Proof Sketch.\/}]\ignorespaces}{\hfill$\blacksquare$\endtrivlist}



\section{Introduction}


In contrast to the simple clausal form, i.e. a conjunction of clauses,   
 the  non-clausal (NC) form, on which  focuses  this article,
allows an arbitrary nesting of the $ \wedge$ and $\vee $ connectives. Thus, 
 the  NC formulas of a given logic contain an undetermined number of nested
  $ \wedge$ and $\vee $ connectives and their atoms 
  are negated and unnegated literals of the given logic.
The   expressiveness of NC formulas  
is exponentially  richer  than that of clausal formulas  
and  they have  found much use in heterogeneous fields and  practical settings as discussed below. 
 
  \vspace{.05cm}
Indeed, within classical logic, non-clausal formulas   are  found in numerous scenarios 
 and reasoning problems such as quantified boolean formulas \cite{DBLP:journals/constraints/EglySW09}, DPLL \cite{ThiffaultBacchusWalsh04}, 
 nested logic programming   \cite{DBLP:journals/tplp/PearceTW09},  knowledge compilation
\cite{DBLP:journals/jacm/Darwiche01}, 
description logics \cite{DBLP:journals/jar/KlarmanES11},      
 numeric planning \cite{ScalaHTR20} and many other 
 fields that are  mentioned in \cite{imaz2021horn}.  In the particular
 case of first-order
 logic, one can find  approaches on non-clausal theorem proving
  in the former steps  of automated reasoning
 e.g., \cite{Bledsoe77,Andrews81} but    such area is  still the object of current research activity as    the     
 regularly reported novel results  show e.g.,
 \cite{FarberKaliszyk19,00010HZC18,OliverO20}. 
 
  \vspace{.05cm}
 And within non-classical logics, non-clausal formulas having different roles and functionalities  have  
  been studied in a profusion of  languages:  
   signed many-valued logic  
 \cite{ MurrayRosenthal94,  BeckertHE98, Stachniak01}, 
 \L ukasiewicz  logic \cite{Lehmke96}, 
 Levesque's three-valued  logic \cite{CadoliS96},   Belnap's four-valued logic 
 \cite{CadoliS96}, 
 M3  logic \cite{Aguilera97},  fuzzy logic \cite{Habiballa12}, fuzzy description  logic \cite{Habiballa07}, 
   intuitionistic logic \cite{Otten17},
  modal logic  \cite{Otten17},  lattice-valued  logic \cite{00010HZC18} and more.

 \vspace{.05cm}
 In many frameworks of   non-classical and classical logics, 
  non-clausal formulas 
  are often translated into clausal form e.g. \cite{Hahnle94a, BaazF95,CadoliS96}
   to which clausal
  reasoning methods are then applied.  However, it is well-known that such translations 
  can either  blow   up   exponentially the size of  formulas or  lose, both,
  their    semantical properties,  preventing its application in some settings,
   and  original syntactical structure,  proven experimentally 
     to  highly decrease practical efficiency.
On the other hand,  the syntactic form of the formulas involved  plays a role
 \cite{CadoliS96}: reasoning in Levesque's three-valued system \cite{Levesque89} is polynomial if the formulas are in clausal form, while it is co-NP-complete if no normal form is assumed.

\vspace{.05cm}
On the other side,     {\it  Horn clausal  formulas}    can be read naturally as 
 instructions for a computer, and  are recognized as central for  
deductive databases, declarative programming, and more generally, for rule-based systems.
In fact, Horn formulas have received  a great deal of attention   
since 1943 \cite{McKinsey43,Horn51}  and, at present, there is a broad span of 
areas within artificial intelligence relying on them, and  their scope covers a fairly large spectrum 
 of   realms spread across many logics and a variety of reasoning settings.

\vspace{.05cm}
Furthermore, regarding Horn efficiency, the valuable contribution  of    the conjunction 
of Horn  formulas and Horn-SAT algorithms to  clausal efficiency 
 is reflected  by the fact that  the highly efficient  DPLL solvers embed a  
 Horn-SAT-like algorithm, so-called   {\em Unit Propagation}\footnote{So the terms 
    {\em Horn-SAT  algorithm} and   {\em Unit Propagation
    procedure}  will be used interchangeably.}\cite{DixonGP04}. 
   These   algorithms have  been greatly optimized to the point that, the Horn-SAT algorithm 
devised for  propositional logic    is even strictly linear 
\cite{DowlingGallier84, DuboisAndreBoufkhadCarlier96}.
    Hence,  searching  for polynomial  (clausal) 
    super-classes of the Horn class in propositional logic has been  
a key issue for several decades in the quest for improving clausal reasoning, and indeed,
currently the existence of  a great number of such classes is known;    
the names of some well-known of these classes are:
hidden-Horn, generalized  Horn, Q-Horn, extended-Horn,   SLUR, 
  Quad,   matched,  UP-Horn and more (see \cite{Prestwich09b,imaz2021horn} for short reviews).

\vspace{.05cm} 
In contrast to such remarkable advances in clausal tractability, the non-clausal tractability 
 is  enormously delayed  
as   the  two following facts clearly reveal: 
 (i)  there is only one (recently found) non-trivial\footnote{Trivial classes include,  for instance,
unsatisfiable formulas 
whose translation to DNF is polynomial.}
  polynomial class in propositional logic \cite{imaz2021horn}; and 
(ii)   beyond propositional logic, there is none of such polynomial non-clausal  classes. 

\vspace{.05cm} 
 Thus, since the signed many-valued logic  is, one may say,  rather close to propositional logic
   and is employed in a wide range of 
 reasoning scenarios and  applications e.g., 
 \cite{BeckertHahnleManya00,LuMurrayRosenthal05,LeitschFermuller05,AnsoteguiBLM13,Fagin20}  
 (discussed  in related work),   we have selected    its sub-class, called regular  logic, in order to determine a first tractable non-clausal class 
  within  approximate reasoning. 

\vspace{.05cm}
For this purpose,  we  first introduce the  hybrid class  of  
{\em  $\mathcal{R}$egular many-valued $\mathcal{H}$orn-NC  formulas, or $\mathcal{RH}$,}
 resulting from suitably merging both regular classes,     Horn 
  and   NC,  or equivalently,  by  {\em suitably  lifting  the existing  regular Horn  pattern 
\cite{Hahnle96,Hahnle01,Hahnle03} to     NC form.} We then 
 prove that  satisfiability testing $\mathcal{RH}$ is polynomial.

Thus, our first contribution is   carried out as follows. 
By lifting   the regular Horn   pattern \cite{Hahnle96,Hahnle01,Hahnle03}  
 (a regular clausal formula  is   Horn if all its clauses have any number of 
 negative  literals and  at most one  positive literal) to the NC level,
we establish the regular  Horn-NC pattern as the next recursive 
non-clausal restriction: {\em a regular NC formula is 
 Horn-NC  if all its disjunctions have any number of  negative  disjuncts and 
at most one non-negative  Horn-NC disjunct.}  
 Accordingly, $\mathcal{RH}$ is the class of  regular  Horn-NC formulas.  
 Note that $\mathcal{RH}$ naturally includes 
 the regular Horn clausal   class.
 Subsequently,  
 we provide a more fine-grained syntactical definition of  $\mathcal{RH}$ in a compact and  inductive function.

\vspace{.05cm}
Our second contribution is proving the relationships between  $\mathcal{RH}$ 
and the regular Horn  and  NC classes    which  are  as follows: (1)  
 $\mathcal{RH}$ is related to the regular Horn class in that
 every Horn-NC  formula  is logically equivalent to a Horn  formula, and 
 hence,   $\mathcal{RH}$ and regular Horn are equivalent  semantically  
but   syntactically $\mathcal{RH}$ subsumes  regular  Horn; 
and (2)   $\mathcal{RH}$  is related to the regular NC class
 in that  $\mathcal{RH}$  contains all regular NCs whose 
 clausal form (to be specified) is Horn.
     The Venn diagram in Fig. 1  relates the  new   $\mathcal{RH}$  to the 
   known   regular classes Horn  ($\mathcal{H}$), Non-Clausal   ($\mathcal{N_C}$) and Clausal  ($\mathcal{C}$).
\vspace{-.1cm}   
\begin{center}
 \begin{tikzpicture}
  \begin{scope}[blend group=soft light]
    \fill[black!40!white]   (270:.9) circle (2.3);
    \fill[green!100!white] (230:1.2) circle (1.3);
    \fill[red!100!white]    (310:1.2) circle (1.3);
\end{scope}
  \node at (100:.8)       {\bf \ $\ \mathcal{N_C}$};
  \node at (215:1.6)      {\bf $\mathcal{C}$};
  \node at (325:1.6)    {\bf $\mathcal{RH}$};
 \node at (270:.9)         {\bf $\mathcal{H}$};
\end{tikzpicture}

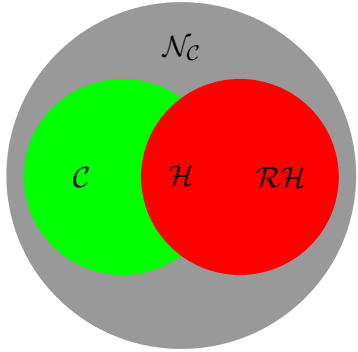
\footnotesize{{\bf Fig. 1.} The Horn$\cdot$NC, Horn, NC  and clausal classes.}
\end{center}

\vspace{0.05cm}
As a third contribution,  we provide the calculus {\em Regular  Non-Clausal   Unit-Resolution}, or 
$\mathit{RUR_{NC}}$,
 prove its  completeness for $\mathcal{RH}$ and  that it enables 
 checking $\mathcal{RH}$ satisfiability      in polynomial time. 
This claim shows that our initial intended aim is achieved
giving that $\mathcal{RH}$ is  multi-valued, non-clausal and tractable, 
and so far as we know,
$\mathcal{RH}$ is {\em the first published class with such features.}
The polynomiality of $\mathcal{RH}$ yields an immediate proof that  the 
computational problem {\em Regular-Horn-NC-SAT is P-complete.}

\vspace{0.05cm}
Some  proven properties \cite{imaz2021horn} of  propositional Horn-NC  formulas  
also apply to  the (?) regular ones presented here,  and  among them,
we highlight  their polynomial recognition, that is, deciding whether any arbitrary 
regular NC formula  is 
 Horn-NC is performed in polynomial time. Altogether, $\mathcal{RH}$ enjoys the advantageous computational properties
of being a class both   recognized and   tested for satisfiability in worst-case polynomial complexity.

\vspace{0.05cm}
We synthesize and illustrate our aforementioned contributions through the  
specific  formula $\varphi  $ given below, whose  infix notation 
is explained in detail in
Section \ref{sec:Non-clausal} and wherein $ \phi_1 $,  $ \phi_2 $ and $ \varphi' $ are 
  regular NCs, and    
$ X^{\geq  \alpha  }$  and $X_{\leq \alpha  }$ denote a  literal that is  satisfiable 
 if the truth-value assigned  to 
 $ X $  is, respectively, {\it greater} or {\it less} than or equal 
  to the  threshold $ \alpha $: 
    $$ \varphi=\{\wedge \ \,  P_{\leq .8  } \ \, (\vee \ \, P_{\leq  .2 }  \ \, \{\wedge \ \, 
(\vee \ \, P_{\leq .3  } \ \, Q_{\leq .4  }  \ \, P^{\geq  1.  }) \ 
\, (\vee \ \, \phi_1 \ \, \{\wedge \ \, \phi_2  \ \,  {P_{\leq .6  } } \}  )
 \ \, Q^{\geq .7  }  \} \,) \ \,  {\varphi'} \}$$
    
\vspace{0.05cm}
We will show that  $ \varphi $ is Horn-NC  when  $ \phi_1 $, $ \phi_2 $
and $ \varphi' $ are Horn-NC  and    at least one of $ \phi_1 $  or  $ \phi_2 $ is negative. In that case we will prove that:

\newpage
1. \  $ \varphi $ can be tested for satisfiability in polynomial time.
 
\vspace{.08cm}
2.  \ $ \varphi $ can be recognized as Horn-NC in polynomial time  \cite{imaz2021horn}.
  
  \vspace{.08cm}
3. \ $ \varphi $ is logically equivalent to a   regular Horn formula.

 \vspace{.08cm}
4. \ $ \varphi $ is exponentially smaller than its equivalent regular Horn formula.
 
 \vspace{.08cm}
5. \    Applying $ \wedge / \vee $ distributivity to $ \varphi $ yields a   regular Horn formula.






\vspace{.15cm}
  Section \ref{sec:logicprogram}   shows  that,
  $\mathcal{RH}$ and $\mathit{RUR_{NC}}$ in tandem
 allow logic programing: $(i)$ 
    enriching their  syntax from simple regular Horn   rules   to regular Horn-NC  rules
 in which   heads and bodies are  NCs with slight syntactical restrictions; and  $(ii)$   
 answering queries with an efficiency   comparable to  clausal efficiency, 
 that is, in polynomial time. This is possible thanks to the facts that
  regular  Horn-NC formulas are, both, polynomial for satisfiability testing and have only
   one minimal model    (indeed, as above mentioned, they are 
   logically equivalent to a regular Horn clausal  formula).
 
 \vspace{.05cm} 
In   future work, we outline how the Horn-NC class and  
    NC Unit-Resolution    will be  defined  in  
     other uncertainty logics\footnote{For the discussion of their extension to classical logics, the 
  reader may consult \cite{imaz2021horn}.}, e.g. {\em  \L ukasiewicz   and possibilistic logics.}
 As  both   entities are basic in  DPLL, they can also be a starting point
towards developing {\em NC DPLL-based  approximate reasoning.}   
Finally, we think that our 
  definition of   NC Unit-Resolution
 is the base to obtain  
  {\em Non-Clausal  Resolution} for  some uncertainty  logics, missing so far.

\vspace{.05cm} 
  The  paper continues as  follows.
  Sections \ref{sec:NCbasis} and  \ref{sec:Non-clausal} present  background 
  on regular clausal and   non-clausal logic, respectively.  Section \ref{sec:definition}    defines 
     $\mathcal{RH}$. Section \ref{sect:HNC-H_NC} relates $\mathcal{RH}$ 
     to the Horn and NC classes. Section  \ref{sec:polynomial}
    introduces  $\mathit{RUR_{NC}}$  and proves the tractability of $\mathcal{RH}$. 
    Section \ref{sec:logicprogram} applies $\mathcal{RH}$ and  $\mathit{RUR_{NC}}$ 
    to  NC logic programming. 
     Sections \ref{sec:relatedwork} and  \ref{sec:futurework} focus on
  related    and  future work, respectively. Last section   
 summarizes our contributions.

\section{Regular Many-Valued Clausal} \label{sec:NCbasis}

      This section   presents  notation,   terminology and  background  on  clausal regular    logic,
      and since regular logic is
a sub-class of  signed logic,  we start by a  general presentation
  of both logics      (for    a  complete presentation, the reader may  
 consult   \cite{BeckertHahnleManya00,Hahnle01,BaazFS01,Hahnle03}).
  

 Signed logic differs from propositional logic only at the literal 
  level. 
  A signed literal
  is a pair $ S \cdot P $, where $  P $ is a proposition and $S  $
  is a (usually finite) set  of  truth-values, and it is 
   satisfiable by an interpretation $ \mathcal{I} $ only if
  $ \mathcal{I} (P) \in  S$.  Since $ \mathcal{I} (P) \in  S$ 
  is true or false, i.e. two-valued, satisfaction of $\wedge /\vee  $-connectives by 
   interpretations is like in classical logic\footnote{See \cite{BeckertHahnleManya99} for a detailed analyze of the relation 
  between signed and propositional logics.}.
  
  \vspace{.05cm}
Signed  logic is a generic
representation for finite-valued  logics since:  deciding the satisfiability
of formulas  of any finite-valued  (and some infinite-valued)  logic
 is  polynomially reducible 
to the problem of deciding the satisfiability of  signed
clausal formulas \cite{Hahnle94a}.

  \vspace{.05cm}
   Regular logic is the most studied sub-class of signed logic
  and derives from it when the  truth-value domain  is  totally ordered and   
  the signs $S$  are intervals 
  of two kinds: 
  $ [- \infty \,, \alpha ]$  or $ [ \alpha \,, \infty ]$.
  The significance of regular logic stems from its close connection 
  with signed logic  \cite{Hahnle96}, that is:
{\em  every signed formula is logically equivalent to some regular formula.}

\vspace{.1cm}   
 The regular language  is the same 
 for  the clausal and the non-clausal frameworks and is defined next.

\begin{definition}  \label{def:alphCNF} The   regular symbol language is formed by  
 an infinite  truth-value set $\mathcal{T} $ endowed with a total ordering $ \succcurlyeq $,
plus the     sets:  regular signs $\{{\geq   }, {\leq   } \}$,  regular propositions 
  $\mathcal{P}=$ \{P, Q, R, \ldots \}, 
classical connectives  \{$\vee,  \wedge\}$ and auxiliary symbols: 
 (,  ),   \{ and   \}. 
\end{definition}


\noindent {\bf Remark}. In the examples throughout the article,
  the truth-value set $\mathcal{T} $  will be the real unit interval  $\mathcal{T}  =[0 , 1]$,  
as  is usually considered in the literature.

\vspace{.2cm}
 Next  we introduce our notation for   atoms.  Although to denote regular literals  a plethora of notations have been invented in the clausal framework, we use a different 
 notation for our non-clausal setting with a twofold purpose: (i) minimizing
the  symbols existing  in regular NC formulas for improving their readability; 
and (ii) helping to visually determine whether a regular
NC formula is Horn-NC.

\begin{definition} If $  \alpha , \beta \in \mathcal{T} $,  
 $ \alpha \geq \beta $   is a    regular constant. If $ X \in \mathcal{P} $ and $  \alpha  \in \mathcal{T}  $, then
  $X^{\geq  \alpha } $  and $ X_{\leq  \alpha } $  are, respectively,   
  a regular  positive and a regular negative literals.      
   $\mathcal{K}$ and $ \mathcal{L}$ are, respectively,  the set of regular constants and regular literals. 
 $ \mathcal{K} \cup \mathcal{L} $ includes  the atoms.
\end{definition} 

\begin{example} $.2 \geq .8 $ is a regular constant; $P^{\geq  .7 }$ and $R_{\leq  .1 }$ are
examples of regular positive and regular negative  literals, respectively.  \qed
\end{example}

 We superscript and subscript positive and negative literals,
respectively, in order to be able to
recognize, just through visual inspection,  whether a
regular NC formula is  Horn-NC.

\begin{definition}
   $C=(\vee \ L_1 \ L_2 \ \ldots \ L_k)$, the $ L_i$   being regular atoms,  is 
a regular clause.  
    $\{\wedge \  C_1 \ C_2 \ \ldots \ C_n\}$, the  $ C_i $ being regular clauses, is a 
regular clausal formula. $ \mathcal{C} $ is the set of regular clausal formulas. 
\end{definition}

\begin{example} \label{ex:clausalexample} $\{ \,\wedge \ (\vee \ \, P^{\geq  .7 } 
 \ \, .2 \geq .8  \ \, R^{\geq 1  }  )
 \ (\vee \ \, P^{\geq  .2 }  \ \,   R_{\leq  0 }  ) \,\}$ is a regular clausal formula. 
\end{example}

\begin{definition} A  regular clause with  at most
one regular positive   literal is regular Horn. If the $ h_i $ are regular Horn clauses then $\{\wedge \  h_1 \ h_2 \  \ldots \ h_n\}$  is a regular Horn formula.
 $ \mathcal{H} $ is the set of regular Horn formulas. 
\end{definition}

\noindent -- The set  $ \mathcal{H} $ is defined in \cite{Hahnle96} and a sub-class of $ \mathcal{H} $ was previously
defined in \cite{EscaladaManya94}.

\begin{example} \label{ex:Hornexample} The formula from Example \ref{ex:clausalexample} 
 is not regular Horn  because  its
 first clause  is not regular Horn, while  
 $ \{\wedge \ (\vee \ \ P^{\geq .7  }  \ \ Q_{\leq .8  }  \ \  R_{\leq .9  }  ) 
\ (\vee \ \ P^{\geq .7  }  \ \  R_{\leq  .1 }  )\} $ is regular Horn. \qed

\end{example}

-- Note that counting the superscripted literals 
occurring in clauses  is enough to recognize whether
 or not a regular clausal formula is  Horn.

\vspace{.05cm}
 {\bf Note.} Since   this article  focuses   on regular logic, 
in most cases  we will omit   the word {\em regular}
preceding  entities and simply speak of literal, clause, formula, etc.

\vspace{.05cm}
-- We will denote {\bf $\top$} any  satisfiable constant, e.g.  $ 1 \geq  .6   $,
and   {\bf $\bot$} any  unsatisfiable  constant, e.g.  $ .6 \geq 1    $, 
whose  formal  definitions  follow.

\begin{definition} \label{def:constant} Let  $ \alpha, \beta \in \mathcal{T} $. We denote {\bf $\top$}  any   constant 
 $  \alpha \geq   \beta  $ such that $ \alpha \succcurlyeq   \beta$, and note
    {\bf $\bot$} any  constant  
  $  \alpha \geq   \beta $ such that    $  \beta \succcurlyeq \alpha $ and $\alpha \neq \beta  $. 
   The   empty conjunction $ \{\wedge\} $ is considered equivalent to
  a {\bf $\top$}-constant and the empty disjunction $ (\vee) $  to a {\bf $\bot$}-constant.
\end{definition}

 \begin{definition} \label{def:interpretation} An interpretation
  $\mathcal{I}$ maps the   propositions $ \mathcal{P} $ into the truth-value set $ \mathcal{T} $ and the 
clausal  formulas $ \mathcal{C} $  
  into $\{0,1\}$ and the mapping
  is extended from  $\mathcal{K} \cup \mathcal{P}$ to  $ \mathcal{C} $   by means of the  rules below,   
where $X \in \mathcal{P}$ and $ \alpha  \in \mathcal{T}$.
\begin{itemize}


\item   $\mathcal{I}(\mathrm{\bot})=\mathcal{I}(\,(\vee)\,)=0$
 \, and \, $\mathcal{I}(\top)=\mathcal{I}(\,\{\wedge\}\,)=1$.

\vspace{-.1cm}
\item   $\mathcal{I}(X^{\geq  \alpha  } )= \left\{
\begin{array}{l l}
1 & \   \mbox{if \ $\mathcal{I}(X) \,\succcurlyeq \, \alpha$}\\
0 & \   \mbox{otherwise}\\
\end{array} \right. 
\: \quad \;  \mathcal{I}(X_{\leq  \alpha  } )= 1 - \mathcal{I}(X^{\geq  \alpha  } )$ 

\vspace{-.1cm}
\item      $\mathcal{I}( \, (\vee\  \ell_1 \ldots \ell_{i}  \ldots  \ell_k) \, ) \,=
\mbox{max}\{\mathcal{I}(\ell_i): 1 \leq i \leq k \}$.

\vspace{-.1cm}
\item       $\mathcal{I}(\, \{\wedge\  C_1 \ldots C_{i}  \ldots   C_k\}\, )
=
\mbox{min}\{\mathcal{I}(C_i): 1 \leq i \leq k \}$.

\end{itemize} 
\end{definition}

\begin{definition} \label{def:model} Some well-known semantical notions  follow, $ \varphi $ being a formula:

\vspace{.2cm}
 -- An interpretation    $\mathcal{I}$ is a   model of   \,$\varphi$  if  \,$\mathcal{I}(\varphi)=1.$ 

\vspace{.05cm}
   -- If $\varphi$  has a model then it is    satisfiable and otherwise unsatisfiable.

\vspace{.05cm}
  --  $\varphi$ and $\varphi'$ are  (logically)  equivalent, noted
 $\varphi \equiv \varphi'$,   if   
  \ $\forall \mathcal{I}$, 
  $\mathcal{I}(\varphi)=\mathcal{I}(\varphi')$.
  
  \vspace{.05cm}
  -- $\varphi'$ is a logical consequence of  $\varphi$, noted
 $\varphi \models \varphi'$,   if   
  \ $\forall \mathcal{I}$, 
  $\mathcal{I}(\varphi)=1 \rightarrow \mathcal{I}(\varphi')=1$.

\end{definition}

\begin{example} For instance, any interpretation $\mathcal{I}$ such that $\mathcal{I}(R)=.1$
satisfies the formula from Example \ref{ex:Hornexample}, which     is hence satisfiable.
\end{example}

\begin{definition}  We  identify the next satisfiability  problems:

\vspace{.2cm}
   --  Reg-SAT is the satisfiability problem of regular  clausal  formulas.

\vspace{.1cm}
  --   Reg-Horn-SAT is Reg-SAT restricted to its  Horn subclass.

\end{definition}

\noindent {\bf Clausal Complexity.}  Reg-SAT is   NP-complete  
\cite{BeckertHahnleManya00,Hahnle01,Hahnle03} and
 Reg-Horn-SAT has complexity $ O(n \log n) $  and $ O(n) $
  for the  infinite-  and  
  finite-valued regular  logics \cite{EscaladaManya94, Hahnle96,Hahnle01,Hahnle03}, respectively.

\section{Regular   Many-Valued Non-Clausal} \label{sec:Non-clausal}
   
   In this section, we    present  regular non-clausal syntactical
   and semantical  concepts\footnote{For    background on propositional non-clausal   concepts, the reader is referred  to    \cite{DBLP:books/daglib/0029942}.} which 
   are quite straightforwardly obtained by generalizing those from the clausal  
   setting.  

\vspace{.15cm}
\noindent  
For the sake of  readability of non-clausal formulas,  we
next justify our chosen notation of them. Thus, we   employ:

\begin{enumerate}
 \item The prefix notation   because it     
       requires only one  $\vee / \wedge$-connective 
      per formula,  while infix notation requires
   $k-1$, $k$  being  the arity of the involved $\vee / \wedge$-connective.
   
  \item  Two symbol  formula 
    delimiters (Definition \ref{def:NCformulas}),     $(\vee  \,\ldots \,)$  for disjunctions and  
 $\{\wedge  \,\ldots \,\}$ for conjunctions, 
  to better 
   distinguish  them   inside
  nested   non-clausal formulas.
  \end{enumerate}

\begin{definition}  \label{def:NCformulas} The  set  $\mathcal{N_C}$ of 
non-clausal formulas\,\footnote{These formulas are also called "negation normal form formulas" in the literature.} is  inductively defined in the usual  way   exclusively from the following rules:
\begin{itemize}
\item   $\mathcal{K} \cup \mathcal{L} \,\subset \,\mathcal{N_C}$.

\item  If  \ $\forall i \in \{1, \ldots k \}$,  $\varphi_i \in \mathcal{N_C}$  \,then    
\,$\{\wedge  \ \varphi_1 \ldots \varphi_{i}  \ldots    \varphi_k\} \in  \mathcal{N_C}$.

\item   If \ $\forall i \in \{1, \ldots k \}$, $\varphi_i \in \mathcal{N_C}$ \,then  \,$\,(\vee  \ \varphi_1  \ldots 
\varphi_{i} \ldots   \varphi_k) \in  \mathcal{N_C}$.
\end{itemize}

\vspace{.1cm}
 -- $\{\wedge    \,\varphi_1 \ldots  \varphi_{i}  \ldots  \varphi_k \}$ and any $\varphi_i$ 
 are called    conjunction and conjunct,  respectively.

\vspace{.1cm}
  -- $(\vee   \,\varphi_1  \ldots  \varphi_{i} \ldots  \varphi_k )$ and any $\varphi_i$ are called  
disjunction  and disjunct, respectively. 

\vspace{.1cm}
-- $\langle \odot  \,\varphi_1  \ldots  \varphi_{i} \ldots  \varphi_k\rangle$    stands for both 
 $(\vee   \,\varphi_1  \ldots  \varphi_{i} \ldots   \varphi_k ) \mbox{\ and \ }
\{\wedge   \,\varphi_1  \ldots  \varphi_{i}  \ldots  \varphi_k \}.$  
\end{definition}

\begin{example} \label{exsec1:varisexamples} Three examples of NC formulas are given below.
We will show that $ \varphi_1 $ is not Horn-NC while $ \varphi_2 $ 
is   Horn-NC. As $ \varphi_3 $ includes $ \varphi_1 $, then $ \varphi_3 $ is not Horn-NC either.

 \vspace{.3cm}
\noindent    $ \bullet \ \varphi_1=\{\wedge  \ \ 
(\vee  \ \ P_{\leq  .4} \ \ Q^{\geq  .7} \ \ .9 \leq .4    \,)
 \ \ (\vee \ \ Q^{\geq .6  }  \  \   \{\wedge \ R_{\leq  .8 }  \ S^{\geq .9  } \ .4 \geq .2 \, \} \, )  \, \}$
 
 \vspace{.3cm}
\noindent    $\bullet \  \varphi_2= (\vee \ \ \{\wedge \ \ P_{\leq .4  } \ \ 1 \geq  0  \} \  \
  \{\wedge  \ \  (\vee  \ \  P_{\leq .3  } \ \ R^{\geq .8  }  \,) \ \ \{\wedge \ \ Q^{\geq  .6 } \ \ (\vee \ \, P^{\geq  .7 } \ \ S_{\leq  .1 }  \,) \,\} \,\}  \,)$  

  \vspace{.3cm}
\noindent   $ \bullet \ \varphi_3=(\vee \ \ \varphi_1 \ \ \{\wedge  \ \ Q^{\geq .6  }   \ \ 
(\vee \ \ \varphi_1 \ \ Q_{\leq .6  }  \ \ \varphi_2) \,\} \ \ 
\{\wedge \ \ \varphi_2 \ \ .9 \geq  .6   \ \    \varphi_1  \,\} \,)$ \qed

\end{example}

\begin{definition} \label{def:sub-for} Subformulas are inductively defined as 
follows.  The unique subformula of an atom ($\mathcal{K} \cup \mathcal{L}   $) is the atom itself, and 
the sub-formulas of a formula 
$\varphi=\langle \odot \  \varphi_1 \ldots \varphi_i \ldots \varphi_k \rangle$ 
  are $\varphi$ itself plus  the sub-formulas of  the     $\varphi_i$'s. 
\end{definition}

\begin{example} The sub-formulas of a clausal formula are the 
formula itself plus its clauses, literals
 and constants.
\end{example}

\begin{definition} \label{def:graphformula}  
NC formulas  are   representable by trees if:  (i) the nodes are:
each atom  is a   {\em   leaf} and
each  occurrence of   
 a $ \wedge /\vee $-connective  is   an {\em  internal node}; and (ii)  {\em the arcs} are:  each
 sub-formula $\langle \odot \  \varphi_1 \ldots  \varphi_{i}  \ldots \,\varphi_k \rangle$    
 is  a  $k$-ary hyper-arc
 linking  the node of  $\odot$      
  with the node of $\varphi_i$, for every $i$,  if $\varphi_i$  is an atom and
 with the node of its connective  otherwise.
\end{definition}

\begin{example}  The graphical representation of  $ \varphi $ from the introduction 
 is given 
in the illustrative  Examples \ref{ex:formulacompl}, in Section \ref{sec:polynomial}. Example
 \ref{ex:cont} provides further examples of DAGs.
\end{example}

\noindent A different,  bi-dimensional   graphical model of NCs
    is handled in    \cite{MurrayRosenthal93,JainBartzisClarke06} and in other works. 
On the other hand, our approach also applies when  non-clausal formulas are represented 
and implemented as directed acyclic graphs (DAGs), which   allow for important savings in both  space and  time.    

\begin{definition} \label{def:DAGs}  An  NC formula $ \varphi $ is modeled by a DAG
if each  sub-formula $ \phi $   is modeled 
by a unique DAG $D_\phi$ and   each $ \phi $-occurrence by a  pointer 
to (the root of) $D_\phi$.
\end{definition}

\begin{example} Let us consider $ \varphi_3 $ from Example \ref{exsec1:varisexamples}. 
 $ \varphi_1 $ and $ \varphi_2 $ should be represented  by    unique 
DAGs, i.e. $ D_{\varphi_1} $ and $ D_{\varphi_2} $, and each  of the two  occurrences 
of both $ \varphi_1$ and $ \varphi_2 $ within $ \varphi_3 $, by a  pointer to their
 corresponding $ D_{\varphi_1} $ or $ D_{\varphi_2} $.
\end{example}

\noindent {\bf  Remark.} Although our approach is also valid for DAGs, for simplicity, we will use  formulas representable by   trees in the illustrative examples throughout this article.


\vspace{.15cm}
\noindent -- In the remaining of this subsection, we present  semantical notions.

\begin{definition}\label{def:IC3} An interpretation
  $\mathcal{I}$ maps the propositions $ \mathcal{P} $ into $ \mathcal{T} $ and  the 
 non-clausal formulas $ \mathcal{N_C} $  
  into $\{0,1\}$ and the mapping
  is extended from $\mathcal{K} \cup \mathcal{P}$ to $ \mathcal{N_C} $   by mappings 
  atoms as done in Definition \ref{def:interpretation} 
  and non-atomic formulas by the next functions:
\begin{enumerate}

\item  [$\bullet$]    $\mathcal{I}( \, (\vee\  \varphi_1 \ldots \varphi_{i}  \ldots  \varphi_k) \, ) \,
\,= \mathrm{max}\{\mathcal{I}(\varphi_i): 1 \leq i \leq k \}$.

\item    [$\bullet$]     $\mathcal{I}(\, \{\wedge\  \varphi_1 \ldots \varphi_{i}  \ldots  \varphi_k\}\, )
\,=\mathrm{min}\{\mathcal{I}(\varphi_i): 1 \leq i \leq k \}$.

\end{enumerate}
\end{definition}

\begin{definition} The concepts  of model, un-satisfiable formula,  logical equivalence 
and logical consequence in Definition \ref{def:model}
are equally defined for non-clausal formulas.
\end{definition}

\begin{example}   
Let us  take $ \varphi_1 $ and $ \varphi_2 $ from Example \ref{exsec1:varisexamples} above:
 (i) any interpretation $ \mathcal{I}  $ s.t.  $ \mathcal{I}(Q)=1  $ 
is a model of $ \varphi_1 $; and (ii)  one can verify that 
$ \varphi_3 \equiv (\vee \ \varphi_1 \ \, \{\wedge \ \, Q^{\geq .6  } \ \varphi_2\})$. 
\end{example}

\begin{definition} Akin to the clausal case, we  identify the next satisfiability  problems:

\vspace{.25cm}
$-$  Reg-NC-SAT is the satisfiability problem of  regular  NC  formulas.  

\vspace{.1cm}
$-$ Reg-Horn-NC-SAT is Reg-NC-SAT restricted to  its Horn-NC subclass.

\end{definition}

 \vspace*{.2cm}
\noindent {\bf Non-Clausal Complexity.}  We can do the next considerations:

\begin{enumerate}
\item [$ \bullet $] Regarding Reg-NC-SAT, one easily verifies that it is NP-complete:
NP-membership  follows straightforwardly since checking whether a given interpretation is a model
of a regular NC formula is trivially done polynomially. NP-completeness follows from:  
Reg-NC-SAT includes Reg-SAT which in turn includes classical SAT.

\item [$ \bullet $] Regarding Reg-Horn-NC-SAT, among our  original results are,
both, {\em the definition of the  regular  Horn-NC class,} $ \mathcal{RH} $, and the proof 
that its  associated satisfiability problem, namely  {\em Reg-Horn-NC-SAT,    is polynomial.}
From this polynomiality, we will trivially prove that Reg-Horn-NC-SAT is P-complete.
\end{enumerate} 

 Next, some simple rules  to simplify formulas  are supplied.

\begin{definition} \label{def:simpl}    
 Constant-free,  equivalent      formulas   are  straightforwardly obtained 
  by applying to  sub-formulas the   simplifying  rules below:  

\vspace{.25cm}
-- Replace \  \,$(\vee \  \top \  \varphi \,)$   \ with \  $   \top $. 

\vspace{.1cm}
--  Replace \ $\{\wedge \   \bot \  \varphi \,\}$  \  with \ $  \bot $.

\vspace{.1cm}
-- Replace \ $\{\wedge \   \top \  \varphi \,\}$   \   with  \ $ \varphi  $.

\vspace{.1cm}
--  Replace \ \,$(\vee \   \bot \  \varphi \,)$  \ \,with  \ $  \varphi$.
 
\end{definition}

\begin{example} \label{ex:constantsfree} The constant-free,   equivalent NC  of 
$ \varphi_2 $  in Example \ref{exsec1:varisexamples} is:
   $$ \varphi=(\vee \ \ P_{\leq  {\tiny .4} }  \ \  \{\wedge \ \ (\vee \ \ P_{\leq  .3 } 
   \ \ R^{\geq .8  }  )  \ \
\{\wedge \ \ Q^{\geq  .6 }  \ \   (\vee \ \  P^{\geq  .7   }  \ \ S_{\leq .1  } ) \, \} \, \}  )$$    

\end{example}

\noindent {\bf Remark.}  For simplicity and since free-constant,    equivalent  formulas are 
 easily obtained,  hereafter we will consider only free-constant formulas.
 
\section{Defining  the Class  $\mathcal{RH}$} \label{sec:definition}


--   HNC is used as a shorthand for Horn-NC.

\vspace{.1cm}
\noindent -- First of all, we need to define   
the negative formulas, which 
     are the  generalization of    negative     literals of the clausal case.
 
 \begin{definition}  \label{def:negative} Negative  formulas
 are  non-clausal formulas having solely negative literals.   $\mathcal{N}^-$ denotes the set of
  negative non-clausal formulas. 
\end{definition}


\begin{example}   Trivially, negative literals are basic negative NC formulas. Another 
example of negative formula is $(\vee \ \ \{\wedge   \  \,P_{\leq 1  }  \  \,R_{\leq  .8 }  \, \} 
\ \ \{\wedge \ \,S_{\leq 1  }  \  \,(\vee \ \,P_{\leq .3  }   \ \,Q_{\leq 0  }  \,)\,\}\,)$. \qed
\end{example}

\noindent -- Next  we  first  define  $\mathcal{RH}$ in a simple way
and then, by taking at closer look, proceed to give its fine-grained definition  
in a compact and inductive function.
 Below we characterize    $\mathcal{RH}$
   by  lifting   the Horn-clausal pattern (defined in 
   \cite{Hahnle96,Hahnle01,Hahnle03} as 
   "a regular Horn clause has at most  one   positive    literal"),
 to the NC level in a straight way  that is as follows.

\begin{definition} \label{theorem:visual} A regular NC disjunction is  HNC 
  if it has at most one 
 disjunct having  positive   literals. A regular NC formula 
 $ \varphi $  is  HNC if all its disjunctions are  HNC.
We denote $\mathcal{RH}$ the class of regular HNC formulas.
\end{definition}

\noindent {\bf Important Remark}.    As  Definition   \ref{theorem:visual} is not concerned with how 
formulas are modeled,
     our approach  also applies when   they  are  represented 
     by  DAGs and not just by trees.

\vspace{.2cm}  
 Clearly regular Horn formulas are regular HNC,   which 
 implies that the published  Horn clausal  class 
 $ \mathcal{H}$ \cite{Hahnle96,BeckertHahnleManya00,Hahnle01,Hahnle03} is
naturally  subsumed by $\mathcal{RH}$, namely $ \mathcal{H} \subset \mathcal{RH}$.
 
  \begin{proposition} \label{prop:allHNFhaveHNF} All sub-formulas of any HNC formula are HNC.
 \end{proposition}

 Such claim follows trivially from Definition \ref{theorem:visual}.
  The converse does not hold as there are non-HNC formulas
whose all sub-formulas are HNC.

\begin{example} \label{exam:simple}  
One can see that $\varphi_1$ below has only one non-negative disjunct and so 
 $\varphi_1$ is HNC,  while $\varphi_2$  is not HNC as it has two non-negative disjuncts.
 \begin{itemize}
\item   $\varphi_1 = (\vee  \   \  \{\wedge \  \ Q_{\leq .6  }  \ \  S_{\leq .7  } \} \  \ 
\{\wedge \  \ R^{\geq  .7 }  \   \ P^{\geq .3  }  \,\} \,).$

\item   $\varphi_2 = (\vee  \ \ \{\wedge  \ \  Q_{\leq  .6 }  \ \ S^{\geq .7  }  \} \ \ 
\{\wedge \ \ R^{\geq .7  } \ \ P_{\leq .3  }  \,\} \,)$. \qed

\end{itemize} 
\end{example} 
 
 Thus superscripting and subscripting     positive and negative literals, respectively,  
   enables   to check how many disjuncts in a given disjunction contain   positive literals,
and so to  decide, according to Definition \ref{theorem:visual}, whether a given regular NC is HNC. 
 
 \begin{example} \label{Ex:morecomplex} We  now consider     
  both formulas $\varphi$  in Example \ref{ex:constantsfree} 
   and $\varphi'$ below, which results from $\varphi$ by just
switching  its  literal $P_{\leq .4  } $  for $P^{\geq .4  }  $:
$$\varphi'= (\vee \ \ P^{\geq  {\tiny .4} }  \ \  \{\wedge \ \ (\vee \ \ P_{\leq  .3 } 
   \ \ R^{\geq .8  }  )  \ \
\{\wedge \ \ Q^{\geq  .6 }  \ \   (\vee \ \  P^{\geq  .7   }  \ \ S_{\leq .1  } ) \, \} \, \}  )$$

\noindent    
Now we only check  whether $\varphi$ and $\varphi'$ are  HNC, and 
later, {\em they will be   thoroughly  analyzed.} So, all     disjunctions of 
$\varphi$, which are $(\vee  \ \, P_{\leq  .3 }  \ \, R^{\geq .8  }  )$, 
$(\vee \  P^{\geq .7  }   \  S_{\leq  .1 } )$ and   $\varphi$ itself,\footnote{Definition \ref{def:sub-for} stipulates
that the sub-formulas of  $\varphi$   include $\varphi$ itself.} have exactly one non-negative disjunct; 
thus $\varphi$  is HNC.
Yet, $\varphi'$ is    of the kind $(\vee \  P^{\geq {\tiny .4}  }  \ \phi)$,
 $\phi$  being  non-negative.  As  $\varphi'$ has two
non-negative disjuncts,   $\varphi'$ is not HNC.  \qed
\end{example}


\noindent Towards a fine-grained definition of $\mathcal{RH}$,  we individually  and inductively specify:

\vspace{.2cm}
$-$          $\mathrm{{HNC}}$ conjunctions,    in Lemma \ref{def:HNCconjunc}, and

\vspace{.1cm}
$-$     $\mathrm{{HNC}}$ disjunctions,   in Lemma \ref{def:disjunHNC},

 \vspace{.2cm}
\noindent  and subsequently,   we {\em compactly}  specify   $\mathcal{RH}$   
 by  merging  the  precedent specifications    into an inductive function  
 given in Definition \ref{def:syntacticalNC}.

\vspace{.15cm}
Clearly, conjunctions of Horn  clausal  formulas
      are    Horn too, 
 and   a similar kind of {\em Horn-like compliance} also holds in NC,   viz.  
    conjunctions of HNC formulas are     HNC too, which is straightforwardly formalized next.
  
  \begin{lemma} \label{def:HNCconjunc} 
  Conjunctions of   
      HNC formulas  are   HNC as well, formally:
  $$\{\wedge \  \varphi_1   \,\ldots\,  \varphi_{i} \ldots \,\varphi_k\} \in \mathcal{RH}
  \mbox{\em \ \ iff \ \ for }   1 \leq i   \leq  k, \ \varphi_i  \in \mathcal{RH}.$$
  \end{lemma}
  
\begin{niceproof} It is obvious that if all  sub-formulas $\varphi_{i}$ individually verify Definition \ref{theorem:visual}
so does a conjunction thereof, and vice versa.
\end{niceproof}

\begin{example}  \label{ex:Semant-Conjunt} If  $H_1$ and $H_2$ are  Horn 
and $\varphi$ is from Example \ref{ex:constantsfree}, 
which by Example \ref{Ex:morecomplex} is HNC, then for instance  
$\varphi_1=\{\wedge \ H_1 \ H_2 \    \varphi  \}$ is HNC. \qed
\end{example}

In order to formally  define  $ \mathcal{RH} $,  we now take a closer look at  Definition \ref{theorem:visual}. Thus,  
   it is not hard to check 
  that the definition of HNC disjunction of Definition \ref{theorem:visual} can be 
  equivalently   reformulated in the next inductive manner:
 {\em "a disjunctive NC is HNC if it has any number of negative disjuncts and  
 at most  one non-negative HNC disjunct",}
 which leads  to the next
    formalization and  statement.

\begin{lemma} \label{def:disjunHNC}  
A disjunctive NC  
$\varphi=(\vee \ \varphi_1 \ldots \varphi_i \ldots  \varphi_k)$ with  $k \geq    1$ disjuncts 
belongs to $ \mathcal{RH} $  iff  
$ \varphi $ has one HNC disjunct and $k-1$  negative   disjuncts,    
 formally 
$$\varphi=(\vee \ \varphi_1 \ldots \varphi_i \ldots     \varphi_k) \in \mathcal{RH} \mbox{\em \ \ iff}$$
  $$\mbox{\em there is} \ i \ \mbox{\em s.t.} \ \varphi_i \in \mathcal{RH} 
\  \mbox{\em and} \ \mbox{\em for all }  j \neq i,  \varphi_j \in \mathcal{N}^-.$$ 
\end{lemma}

\begin{niceproof} {\bf If:}
 Since    the sub-formulas 
$\forall j, j \neq i,  \varphi_j$
have no positive literals, the non-negative disjunctions
of $\varphi= (\vee \ \varphi_1 \ldots  \varphi_i \ldots     \varphi_k) $  
 are those of $ \varphi_i $ plus  $ \varphi_i $ and $ \varphi $ themselves.  
Given that  by hypothesis $ \varphi_i \in \mathcal{RH}  $ and that 
$ \forall j, j \neq i, \varphi_j $ has no positive literals then  clearly
all of them belong to $ \mathcal{RH}$. Furthermore, since $ \varphi$ has only 
one non-negative disjunct and its sub-formulas verify Definition \ref{def:NCformulas}, 
 so does $ \varphi $ itself.
\noindent {\bf Iff:} It is proven by contradiction (a similar proof is given in the first 
theorem  in the Appendix):  if  
  any of the  two conditions of the lemma does not hold,
 i.e. (i) $ \exists i, \varphi_i \notin \mathcal{RH} $  or (ii)
 $\exists i,j, i \neq j,  \varphi_i,\varphi_j \notin \mathcal{N}^-$,  then 
$  \varphi \notin \mathcal{RH}$. 
%
\end{niceproof}

\noindent The next claims follow trivially from Lemma \ref{def:disjunHNC}:
 
 \vspace{.2cm}           
  $\bullet$  Horn clauses are  non-recursive HNC disjunctions.
  
  \vspace{.1cm}        
  $\bullet$ NC disjunctions with all negative disjuncts are HNC.
 
 \vspace{.1cm}
  $\bullet$  NC disjunctions with $ k {\geq   } 2$      non-negative disjuncts are not HNC.  
 
\vspace{.2cm}
\noindent Next, we first reexamine, bearing Lemma \ref{def:disjunHNC} in mind,
 the  formulas from Example  \ref{exam:simple}, included in  Example \ref{exam:NF},  
 and  then those from Example \ref{Ex:morecomplex}, included in Example \ref{ex:disjunHNC}.

\begin{example} \label{exam:NF} Below we analyze   $\varphi_1$  and $\varphi_2$  from  Example \ref{exam:simple}.

\begin{itemize}

\item    $\varphi_1 = (\vee  \   \  \{\wedge \  \ Q_{\leq .6  }  \ \  S_{\leq .7  } \} \  \ 
\{\wedge \  \ R^{\geq  .7 }  \   \ P^{\geq .3  }  \,\} \,).$  

-- Clearly $\{\wedge \  \ Q_{\leq  .6 }  \ \  S_{\leq .7  } \} \in \mathcal{N}^-$.   

-- By Lemma \ref{def:HNCconjunc},
$\{\wedge \  \ R^{\geq  .7 }  \   \ P^{\geq .3  }  \,\} \in \mathcal{RH}$.

-- According to Lemma \ref{def:disjunHNC},  $\varphi_1 \in \mathcal{RH}$.

\item   $\varphi_2 = (\vee   \   \   \{\wedge  \ \  Q_{\leq  .6 }  \   \ S^{\geq .7  }  \} \  \ 
\{\wedge \  \ R^{\geq  .7 }  \  \ P_{\leq  .3 }  \,\} \,)$.

-- Obviously $\{\wedge  \ \  Q_{\leq  .6 }  \   \ S^{\geq .7  }  \} \notin \mathcal{N}^-$ 
and  $\{\wedge \  \ R^{\geq .7  }  \  \ P_{\leq  .3 }  \,\} \notin \mathcal{N}^-$.

-- According to Lemma \ref{def:disjunHNC}, 
  $\varphi_2 \notin \mathcal{RH}$. \qed

\end{itemize}
    
\end{example}

\begin{example} \label{ex:disjunHNC} Consider again     $\varphi$ 
from Example \ref{ex:constantsfree} and $\varphi'$  from Example \ref{Ex:morecomplex}  and 
     recall that $\varphi'$ results from $\varphi$ by  
   just  switching  its literal  $P_{\leq .4  } $ for $P^{\geq .4  } $.   
%
%
 Below we     analyze  one-by-one
 the sub-formulas of both  $\varphi$ and  $\varphi'$. 
\begin{itemize}

\item By  Lemma \ref{def:disjunHNC}, $(\vee  \ \ P_{\leq .3  }  \ \ R^{\geq .8  }  ) \in \mathcal{RH}$.

\item By  Lemma \ref{def:disjunHNC},  $(\vee \ \ P^{\geq .7  }  \  \  S_{\leq .1  } )  \in \mathcal{RH}$.

\item  By  Lemma  \ref{def:HNCconjunc}, 
 $\{\wedge \ \ Q^{\geq  .6 }  \ \   (\vee \  \ P^{\geq .7  }  \ \  S_{\leq  .1 } ) \, \}\in \mathcal{RH}$. 
 
\item  By  Lemma  \ref{def:HNCconjunc}, 
 $\phi=\{\wedge  \  \ (\vee  \ \ P_{\leq .3  }  \ \ R^{\geq  .8 }  )  \ \ 
\{\wedge \ \ Q^{\geq .6  }  \ \   (\vee \ \ P^{\geq  .7 }  \ \  S_{\leq  .1  }) \, \} \, \} \in \mathcal{RH}$.
 

\item Using previous formula $\phi$, we have $\varphi= (\vee \ P_{\leq .4  }   \ \,\phi \,)$. 

 \hspace{.5cm} -- Since $P_{\leq   .4 }  \in \mathcal{N}^-$
and   $\phi \in \mathcal{RH}$, by Lemma \ref{def:disjunHNC}, $\varphi \in \mathcal{RH}$.
    
\item  The second formula in  Example \ref{Ex:morecomplex}  is 
$\varphi'=(\vee \   P^{\geq .4  }  \ \phi \,)$. 

\hspace{.5cm} -- Since $P^{\geq   .4 } ,\,\phi \notin \mathcal{N}^-$,   by Lemma \ref{def:disjunHNC},  $\varphi' \notin \mathcal{RH}$. \qed
  
\end{itemize}
\end{example}

 $ - $ By using   Lemmas  \ref{def:HNCconjunc}  and    \ref{def:disjunHNC}, the class  $\mathcal{RH}$ is syntactically, compactly and inductively defined as follows.

\begin{definition} \label{def:syntacticalNC}   
We inductively define the set   of   formulas  $\mathcal{\overline{HR}}$  from exclusively
    the  rules below,  wherein
 $k \geq    1$ and $\mathcal{L}$ is the set of   literals.
\begin{itemize}

\item [(1)] \ $\mathcal{L} \subset \mathcal{\overline{RH}}.$    \hspace{10.05cm}   

\item [(2)]  \  If  \ $\forall i, \,\varphi_i  \in \mathcal{\overline{RH}} \ \  \mbox{then} \  \
   \{\wedge \  \varphi_1   \,\ldots\, \varphi_{i} \ldots  \varphi_k\} 
   \in \mathcal{\overline{RH}}.$ \hspace{3.27cm}  

\item  [(3)]  \  If \  $\varphi_i \in \mathcal{\overline{RH}}$ \  and \  
 $\forall j \neq i$,  $\varphi_j \in \mathcal{N}^- \  \ \mbox{then} \ \ (\vee \ \varphi_1 \ldots \varphi_i \ldots   \,\varphi_k) 
 \in \mathcal{\overline{RH}}.$ \hspace{.24cm}  
 
\end{itemize}

\end{definition}

\noindent  We prove next that the class $\mathcal{\overline{RH}}$ 
indeed coincides  with the class  $\mathcal{{HR}}$, namely  Definition \ref{def:syntacticalNC}  
is indeed the detailed, recursive and compact definition of $\mathcal{{HR}}$.

\begin{theorem} \label{th:HNFequality}  We have: \ $\mathcal{\overline{RH}}=\mathcal{RH}$.
\end{theorem}

\begin{niceproof}  We prove first $\mathcal{\overline{RH}} \subseteq \mathcal{RH}$ and then 
$\mathcal{\overline{RH}} \supseteq \mathcal{RH}$.

\vspace{.05cm}
 $\bullet$ $\mathcal{\overline{RH}} \subseteq \mathcal{RH}$
    is easily proven by structural induction as outlined below:
   
   \vspace{.05cm} 
   ({1})   $\mathcal{L} \subset \mathcal{RH}$  trivially holds. 
 
\vspace{.05cm} 
 ({2}) The non-recursive $ \mathcal{\overline{RH}} $ conjunctions  are  conjunctions of literals, which trivially verify
 Definition  \ref{theorem:visual} and so are in $\mathcal{RH}$. Further, 
  assuming that $\mathcal{\overline{RH}} \subseteq \mathcal{RH}$ holds 
  until  a given inductive step and that  $\varphi_{i} \in \mathcal{\overline{RH}}, 1 {\leq   } i {\leq   } k$, in the next induction step any formula $\varphi= \{\wedge \  \varphi_1   \,\ldots\, \varphi_{i} \ldots  \varphi_k\} $
 may be added to $ \mathcal{\overline{RH}} $; but  by  Lemma \ref{def:HNCconjunc}, 
 $ \varphi \in \mathcal{RH}$ and  so $\mathcal{\overline{RH}} \subseteq \mathcal{RH}$  holds. 

\vspace{.05cm}
 ({3}) The non-recursive    disjunctions in $ \mathcal{\overline{RH}} $  are obviously  Horn clauses, which trivially 
fulfill Definition \ref{theorem:visual} and so are in $\mathcal{RH}$. Then
assuming that for a given recursive level  $\mathcal{\overline{RH}} \subseteq \mathcal{RH}$ holds,
in the next recursion, only  disjunctions $ \varphi $  in ({3}) are added to $\mathcal{\overline{RH}}$. 
But  the  condition of ({3})  and that of 
 Lemma \ref{def:disjunHNC} are equal; so by  Lemma \ref{def:disjunHNC},  
 $ \varphi $  is in  $\mathcal{RH}$ also.  Therefore $\mathcal{\overline{RH}} \subseteq \mathcal{RH}$ holds.
 
 \vspace{.05cm} 
 $\bullet$ $\mathcal{RH} \subseteq \mathcal{\overline{RH}}$. 
Given  that the structures to define $\mathcal{{N}_{C}}$ and $\mathcal{\overline{RH}}$ 
in Definition \ref{def:NCformulas}  and Definition \ref{def:syntacticalNC}, respectively,    
   are equal, the potential inclusion  of each  NC  
   formula $ \varphi$ in   $\mathcal{\overline{RH}}$ is systematically considered. 
 Further, the statement:
  if $ \varphi \in  \mathcal{RH}$ then $ \varphi \in  \mathcal{\overline{RH}}$, 
    is    proven by structural induction on the depth of formulas, 
     by applying a  reasoning   similar to that of 
  the previous $\mathcal{\overline{RH}} \subseteq \mathcal{RH}$ case and by also using  Lemmas \ref{def:HNCconjunc}
  and  \ref{def:disjunHNC}.
\end{niceproof}

Within the propositional logic setting, the homologue  of  Definition  \ref{def:syntacticalNC} 
has served in \cite{imaz2021horn}
to design   a  linear algorithm  that decides whether a given 
     NC $ \varphi $ is HNC. This algorithm is extensible 
     to regular logic albeit  its polynomial degree can of course slightly increase.

\begin{example} \label{ex:ExamComplet} We analyze   $\varphi$ and $\varphi'$ from Example \ref{ex:disjunHNC} w.r.t. Definition \ref{def:syntacticalNC}: 
\begin{itemize}
\item By   ({3}),   $(\vee  \ \ P_{\leq  .3 }  \ \ R^{\geq .8  }  ) \in \mathcal{{RH}}$. 

\item By   ({3}), $(\vee \ \ P^{\geq .7  }  \ \  S_{\leq .1  } )  \in \mathcal{{RH}}$. 

\item  By    ({2}),  $\{\wedge \ \ Q^{\geq .6  }  \ \ (\vee \  P^{\geq .7  } \ \  S_{\leq  .1 } ) \, \}
 \in \mathcal{{RH}}$. 

\item  By   ({2}), 
 $\phi=\{\wedge  \  \ (\vee  \ \ P_{\leq .3  }  \ \ R^{\geq .8  }  )  \ \ 
\{\wedge \ \ Q^{\geq .6  }  \ \   (\vee \ \ P^{\geq  .7 } \ \  S_{\leq  .1 } ) \, \} \, \} 
\in \mathcal{{RH}}$.
 
\item By  ({3}), $\varphi= (\vee \ \ P_{\leq .4  } \ \ \phi \,) \in \mathcal{{RH}}$

\item  By   ({3}),
$\varphi'=(\vee \ \ P^{\geq  .4 } \ \ \phi \,) \notin \mathcal{{RH}}$. \qed

\end{itemize}
\end{example}

\begin{example} \label{ex:H*NF} If we assume that 
 $\varphi_1$, $\varphi_2$ and  $\varphi_3$ are negative  and 
 $\varphi_4$ and $\varphi_5$ are  HNC, then  according to Definition \ref{def:syntacticalNC},   
 four examples of  {\em nested}  HNC formulas  follow. 
\begin{itemize}

 \item  By    ({3}), \ $\varphi_6=(\vee \ \  \varphi_1 \ \  \varphi_4 ) \in \mathcal{{RH}}$.

 \item  By    ({2}), \ $\varphi_7=\{\wedge \ \  \varphi_1 \ \  \varphi_5 \ \ \varphi_6 \} 
 \in \mathcal{{RH}}$.

 \item  By    ({3}), \ $\varphi_8=(\vee \ \ \varphi_1 \  \ \varphi_2 \ \  
   \varphi_7) \in \mathcal{{RH}}$.
   
\item By  ({2}), \ $\varphi_9=\{\wedge \ \  \varphi_6 \ \  \varphi_7 \ \ \varphi_8 \} 
 \in \mathcal{{RH}}$. \qed

\end{itemize}

\end{example}

\noindent Next, we analyze a more complete example, concretely the one given in the Introduction.

\begin{example} \label{ex:introduction} Let us take $ \varphi $ below, wherein 
$ \phi_1, \phi_2$ and  $\varphi'$ are NC formulas:

\vspace{.25cm}
$\varphi=\{\wedge \ \,  P_{\leq .8  }  \ \, $ 

\vspace{.15cm} 
\hspace{1.5cm}$ (\vee \ \ P_{\leq  .2 }  \ \ \{\wedge \ \ 
(\vee \ \ P_{\leq .3  } \ \ Q_{\leq .4  }  \ \ P^{\geq  1.  }) \ 
\ (\vee \ \ \phi_1 \ \ \{\wedge \ \ \phi_2  \ \  {P_{\leq .6  } } \}  )
 \ \, Q^{\geq .7  }  \} \,)$

\vspace{.15cm}  
    \hspace{2.2cm} $  {\varphi'} \ \}.$

\vspace{.25cm}
\noindent The  disjunctions of $ \varphi $ and the proper $ \varphi $ can be rewritten as follows: 
\begin{itemize}
\item $\omega_1= (\vee \  P_{\leq .3  } \  Q_{\leq .4  }  \  P^{\geq  1.  }) $.

\item   $\omega_2= (\vee \ \phi_1 \  \{\wedge \ \phi_2  \  {P_{\leq .6  } } \} \, ) $.

\item  $\omega_3=(\vee \ \ P_{\leq  .2 }  \ \ \{\wedge \ \ \omega_1 \ \ 
\omega_2 \ \, Q^{\geq .7  }  \} \,) .$

\item  $ \varphi = \{\wedge \ \ P_{\leq .8  } \ \ \omega_3 \ \ \varphi' \}.$

\end{itemize}

\noindent  We analyze one-by-one such disjunctions and finally the proper $ \varphi $:
\begin{itemize}

\item  $ \omega_1$: \ Trivially, $ \omega_1 $ is Horn. 

\item  $ \omega_2 $: \
$ \omega_2 \in \mathcal{{RH}}$   \  if \ $ \phi_1, \phi_2 \in \mathcal{{RH}} $ \ and also if 
at  least one of  
\ $ \phi_1$  or $ \phi_2$ is negative.

\item  $ \omega_3 $: \ $ \omega_3 \in \mathcal{{RH}}$  
\ if  \   $\omega_2 \in \mathcal{{RH}}$ (as $ \omega_1 \in \mathcal{{RH}}$).
 
\item  $ \varphi $: \   
 $ \varphi \in \mathcal{{RH}}$  \ if \ $ \omega_2, \varphi' \in \mathcal{{RH}}$
 (as $ \omega_3 \in \mathcal{{RH}}$  
if     $\omega_2 \in \mathcal{{RH}}$).
\end{itemize}

\noindent Summarizing the conditions  on   $ \varphi $ and on $ \omega_2 $, we have that:
 
 \vspace{.2cm} 
$\bullet$ $ \varphi $ is HNC:  \underline{if}  $\varphi'$, $\phi_1$ and $ \phi_2$   are  HNC
and  \underline{if at least one of} $ \phi_1 $   or $ \phi_2 $ is negative.

\vspace{.2cm} 
\noindent Since 
by Proposition \ref{prop:allHNFhaveHNF}, all sub-formulas of an HNC must be HNC, we 
can conclude that $ \varphi $ is HNC if its sub-formulas are HNC and if 
at least one of $ \phi_1 $   or $ \phi_2 $ is negative. \qed

\end{example}

\section{Relating $ \mathcal{RH}$ to the Classes Horn and NC} \label{sect:HNC-H_NC}
  
  In this section,  we demonstrate the relationships between $ \mathcal{RH}$ and
  the  classes  $ \mathcal{H}$  and  $ \mathcal{N_C}$. 
 The  formal proofs of the theorems were   
     given in \cite{imaz2021horn} but   are provided in an Appendix
   for the sake of the paper be self contained.
 
 \vspace{.15cm}
      A new key and simple concept  is introduced next,  
  necessary to  provide  afterwards, the  relationship between  $ \mathcal{RH}$ and the classes 
      $ \mathcal{H}$ and $ \mathcal{N_C}$.

\begin{definition} \label{def:ECNF}  For every $\varphi \in \mathcal{N_C}$,  we define   $cl(\varphi)$ as
    the unique clausal formula    that
  results from   applying  the
 $\vee/\wedge$  distributivity laws to $\varphi$ until a clausal formula, viz. $cl(\varphi)$, is obtained.
 We will  call $cl(\varphi)$ the clausal form of $\varphi$.
\end{definition}

\begin{example} \label{exa:ECNF} 
Applying $\vee/\wedge$ distributivity to $\varphi_1$ and $\varphi_2$ in  Example \ref{exam:simple},  
 one obtains  the clausal forms  below.
Note  that  $cl(\varphi_1)$ is   Horn  but not    $cl(\varphi_2)$. 
\begin{itemize}
\item  $ \ cl(\varphi_1)=\{\wedge \  \,  (\vee \ \, Q_{\leq .6}   \  \,  R^{\geq .7}) \  \,   
 (\vee \ \, Q_{\leq .6} \ \, P^{\geq .3}) \ \, (\vee \ \, S_{\leq .7} \ \, R^{\geq .7}) \ \,
 (\vee \ \, S_{\leq .7} \ \, P^{\geq .3}) \,\}$ 
 
\item $ \ cl(\varphi_2)=\{\wedge \  \,  (\vee \ \, Q_{\leq .6}   \  \,  R^{\geq .7}) \  \,   
 (\vee \ \, Q_{\leq .6} \ \, P_{\leq .3}) \ \, (\vee \ \, S^{\geq .7} \ \, R^{\geq .7}) \ \,
 (\vee \ \, S^{\geq .7} \ \, P_{\leq .3}) \,\}$  \qed
 
\end{itemize}

\end{example}

 Obviously the   distributivity laws cause the exponential blowup of   $cl(\varphi)$.

\begin{proposition} \label{prop:secondprop} We have that:  $\varphi \equiv cl(\varphi)$.
\end{proposition}
  
  -- The proof follows  from the fact that 
  $cl(\varphi)$  results  by just applying the $\vee/\wedge$  distributivity laws to  $ \varphi $
  and that such laws of course preserve the logical equivalence.
  
  \vspace{.15cm} 
  -- $cl(\varphi)$ is key 
  to relate  $\mathcal{RH}$ to the classes  $ \mathcal{H}$ and $ \mathcal{N_C}$ 
  as   the next statements will show.
  
\begin{theorem}  \label{theorem:first} 
The clausal form of   all HNC formulas is Horn, formally: 
$$\forall \varphi \in \mathcal{RH}, \ \mathrm{we \ have:} \ cl(\varphi) \in \mathcal{H}.$$
\end{theorem}

\begin{niceproof}  See Appendix.
\end{niceproof}

  Theorem \ref{theorem:first} and Proposition \ref{prop:secondprop} yield the
 next characterization  of $ \mathcal{RH}$.

\begin{corollary} \label{cor:first}
Every HNC formula   is logically equivalent to some Horn formula, formally 
$$ \forall \varphi \in  \mathcal{RH}, \,\exists H \in \mathcal{H} \mbox{ such that } \varphi \equiv H.$$
\end{corollary}

\begin{niceproof} From Proposition \ref{prop:secondprop} and Theorem \ref{theorem:first}, 
we have that, for  every HNC formula: $ \forall\varphi \in \mathcal{RH}$, both, 
 $ cl(\varphi) \in   \mathcal{H}$ and
  $ \varphi \equiv cl(\varphi)$.  Hence the corollary holds.
\end{niceproof}

Taking into account the previous corollary and the fact that 
$ \mathcal{H} \subset \mathcal{RH} $, we verify that: 

\begin{corollary} \label{cor:equiva-classes}  
$ \mathcal{RH}$ and $ \mathcal{H}$ are logically equivalent, noted $ \mathcal{RH} \equiv \mathcal{H}$:
every formula in a class is equivalent to another formula in the other class.
\end{corollary}

\begin{niceproof} It follows from Corollary \ref{cor:first} and the fact 
that $ \mathcal{H} \subset \mathcal{RH} $.
\end{niceproof}

$ - $ Therefore, we previously checked that  syntactically $ \mathcal{RH}$ 
subsumes the regular Horn class but now we have proved that semantically
both classes are equivalent.

\vspace{.05cm}
$ - $ Thus $cl(\varphi)$ is a syntactical and semantical means of  
 characterizing $ \mathcal{RH}$; indeed,  because
$cl(\varphi)$  issues from $ \varphi $ by a  syntactical operation,
    and   because $cl(\varphi)$ is semantically equivalent to $ \varphi $, respectively.

\vspace{.05cm}
$ - $ The next theorem specifies which NC formulas  are contained in  $\mathcal{RH}$, or
in other words,  which syntactical NC fragment constitutes $\mathcal{RH}$.

\begin{theorem} \label{prop:sem-NC} 
All  NC formulas  $ \varphi $ whose clausal form is Horn are HNC, namely 
   $$\forall \varphi \in \mathcal{N_C}, \mathrm{\ if \ } cl(\varphi) \in \mathcal{H} \mathrm{\ then \ }  \varphi \in \mathcal{RH}.$$
\end{theorem}

   \begin{niceproof}  See Appendix.
   \end{niceproof}

\begin{example}  \label{ex:HNC} We apply below Theorem \ref{prop:sem-NC} to   Examples \ref{exam:simple}
and   \ref{Ex:morecomplex}.  
\begin{itemize}

\item  {\em Example \ref{exam:simple}:} by Example
\ref{exa:ECNF},    $cl(\varphi_1) \in \mathcal{H}$ and 
 $cl(\varphi_2) \notin \mathcal{H}$;  hence, only $\varphi_1$ is HNC.
 
\vspace{.1cm}
\item   {\em Example \ref{Ex:morecomplex}:}  we do not 
 supply $cl(\varphi)$ nor $cl(\varphi')$    due to their big size,  but
  one  has $cl(\varphi) \in \mathcal{H}$ and   $cl(\varphi') \notin \mathcal{H}$. 
Hence only $ \varphi $ is HNC. \qed
\end{itemize}
\end{example}

 Next Theorem \ref{Theorem:last} just puts together previous
Theorems  \ref{theorem:first} and \ref{prop:sem-NC} and can be viewed as an alternative
definition of  $\mathcal{{RH}}$ to that given in Definition \ref{def:syntacticalNC}.

\begin{theorem} \label{Theorem:last} The next statement holds:
\begin{itemize}
\item    \  
 \ $\forall \varphi \in \mathcal{N_C}:$ \ $\varphi \in \mathcal{{RH}}$ \ {\em iff} \ 
 $cl(\varphi) \in \mathcal{H}$. 
 
\end{itemize}
\end{theorem}

\begin{niceproof} It  follows immediately from Theorems    \ref{theorem:first} and
\ref{prop:sem-NC}.
\end{niceproof}

   Although  the definition of $\mathcal{{RH}}$  in the previous theorem 
   is concise and simple,   trying
   to recognize HNF formulas via $ cl(\varphi) $ is unfeasible, given that 
   obtaining $ cl(\varphi) $ takes both exponential  time and space.  Contrary to this,
   a polynomial algorithm can be obtained following  Definition \ref{def:syntacticalNC} 
   as done in \cite{imaz2021horn} for the propositional logic.

\section{Non-Clausal Unit-Resolution   and the  Tractability of $\mathcal{RH}$} \label{sec:polynomial}

This section   defines 
 {\em   Regular Non-Clausal Unit-Resolution,}  
or $\mathit{RUR_{NC}}$,  which  extends the same rule for propositional logic presented 
in \cite{imaz2021horn}.     $\mathit{RUR_{NC}}$ is  proven to be complete
  for  $\mathcal{RH}$ and to polynomially test $\mathcal{RH}$ for satisfiability.
$\mathit{RUR_{NC}}$ encompasses the main inference rule,  called  RUR, 
and  several simplification rules. The latter rules are simple but   RUR is quite elaborate,
and so we present it    progressively:

\vspace{.15cm}
$-$  for  almost-clausal HNCs, in Subsection \ref{sec:UR-NC};  
 
 \vspace{.05cm}
$-$   for   general HNCs, in Subsection \ref{subsec:generalHNF}; and 

 \vspace{.05cm}  
$-$  for general NCs, in Subsection \ref{subsec:Further-infer}.

 \vspace{.2cm}
\noindent {\bf Remark.} If  $ \varphi $ is a disjunctive HNC
  with more than one disjunct, then, 
 according to Definition  \ref{theorem:visual}, it has at least one disjunct containing
only negative literals and so  assigning 0 to all its propositions satisfies $ \varphi $.
Therefore,  to discard the case in which the input may be trivially satisfiable,
we will consider that \underline{the input $ \varphi $ is a conjunctive HNC formula}.

\subsection{Almost-Clausal HNC formulas} \label{sec:UR-NC}

\vspace{.25cm}
 We start by recalling below  regular   clausal  unit-resolution  \cite{Hahnle96, BeckertHahnleManya00}, wherein   $ X \in \mathcal{P}$ is a
 proposition, $ \alpha, \beta \in \mathcal{T} $  are truth-values and the $ \ell_i $'s are literals:

$$\frac{{\textcolor{red} {X^{\geq \alpha  } }}     \ {\textcolor{red} {\wedge}} \
({\textcolor{blue} {\vee}} \ \ \ell_1 \ \ldots \ell_j \ 
{\textcolor{blue} {X_{\leq  \beta } }} \ \ell_{j+1} \ \ldots \ \ell_k) , \ \alpha > \beta }
{({\textcolor{blue} {\vee}} \ \ \ell_1 \ \ldots \ell_j  
 \ \ell_{j+1} \ \ldots \ \ell_k) }{\mathrm{}}$$

 $ - $ Such rule  is refutationally complete for the regular Horn class \cite{Hahnle96}.

\vspace{.1cm}
 $-$ At first, we introduce {\em RUR}  just for   almost-clausal HNC formulas. 
 Assume  HNCs with the next almost-clausal pattern, 
 where $ X \in \mathcal{P} $ and $ \alpha, \beta \in \mathcal{T} $:   
$$\{{\textcolor{red} {\wedge}}  \ \pi_1 
 \ {\textcolor{red} {X^{\geq   \alpha } }} \  \pi_2 \
({\textcolor{blue} {\vee}} \ \, \phi_1 \, \ldots \, \phi_{j-1} \ 
{\textcolor{blue} {X_{\leq   \beta } }} \ \phi_{j+1} \, \ldots \, \phi_k) \ \pi_3 \}, \  \alpha >  \beta$$

\noindent in which the $ \pi $'s are lists of HNC formulas, namely $ \pi_1=\varphi_1  \, \ldots  \, \varphi_{l-1}$; $\pi_2=\varphi_{l+1}  \, \ldots  \, \varphi_{i-1}  $;
and $ \pi_3= \varphi_{i+1} \, \ldots \, \varphi_n$.
 These formulas are almost-clausal in the sense that 
if the $\varphi$'s and   $\phi$'s were clauses and  literals, respectively, then 
such formulas would  be  clausal.  Since ${\textcolor{red} {X^{\geq   \alpha } }}  $ 
and ${\textcolor{blue} {X_{\leq   \beta } }}$, for $\alpha > \beta$, are unsatisfiable,  
almost-clausal formulas are clearly equivalent to:  
$$\{{\textcolor{red} {\wedge}}  \ \pi_1 
 \ {\textcolor{red} {X^{\geq   \alpha } }} \  \pi_2 \ 
({\textcolor{blue} {\vee}} \  \phi_1 \, \ldots \, \phi_j  
\, \phi_{j+1} \, \ldots \, \phi_k) \ \pi_3 \}$$

\noindent and thus, one obtains   the next   simple   inference rule:  
$$\frac{{\textcolor{red} {X^{\geq   \alpha } }}     \ {\textcolor{red} {\wedge}} \
({\textcolor{blue} {\vee}} \ \ \phi_1 \ \ldots \phi_j \ 
{\textcolor{blue} {X_{\leq   \beta }}} \ \phi_{j+1} \ \ldots \ \phi_k) , \  \alpha >  \beta }
{({\textcolor{blue} {\vee}} \ \ \phi_1 \ \ldots \phi_j  
 \ \phi_{j+1} \ \ldots \ \phi_k) }{\mathrm{\,RUR}}$$

\noindent Note that if  almost-clausal formulas  are    clausal, 
 then the above rule recovers  
regular clausal unit-resolution. By noting  
$\mathcal{D}({\textcolor{blue} {X_{\leq  \beta  } }})$
   the disjunction
$({\textcolor{blue} \vee} \   \phi_1   \ldots   \phi_j \,  \phi_{j+1}   \ldots   \phi_k)$,
 the previous  rule    can be  rewritten concisely as: 
\begin{equation} \label{eq:dsimpleUNC}
\frac{ {\textcolor{red} {X^{\geq  \alpha } }} \ \, {\textcolor{red}\wedge} \ \, 
({\textcolor{blue} {\vee}} \ \ {\textcolor{blue} {X_{\leq   \beta } }}
 \ \ \mathcal{D}({\textcolor{blue} {X_{\leq   \beta } }}) \,), \  \alpha >  \beta}
{  \mathcal{D}({\textcolor{blue} {X_{\leq  \beta  } }})  }{{\mbox{\,RUR}}}
\end{equation}

\begin{example} \label{ex:withoutconj}  Let us consider the next formula:
 $$  \{{\textcolor{red} {\wedge}} \  \, {\textcolor{red} {P^{\geq .8  } }} \ \,    
(\vee   \ \,  Q _{\leq  .4 }  \   {\textcolor{blue} {P_{\leq .5  } }} ) \ \, 
(\vee \ \, R^{\geq  .9 }  \ \,   \{\wedge \ \ Q_{\leq .2  }   \  P_{\leq  .3  } \} )   \,   \}$$

\noindent If we pick up    ${\textcolor{red} {P^{\geq .8  } }}$  and  
 ${\textcolor{blue} {P_{\leq  .5 } }}$, then
we have    
$\mathcal{D}({\textcolor{blue} {P_{\leq  .5 } }})=(\vee \ \,
  Q _{\leq .4  }  )$. 
So by applying the previous rule to $ \varphi $ and then by removing the 
 generated redundant $ \vee $-connective,   
one deduces:

\vspace{.3cm}
\hspace{3.cm} $  \{{\textcolor{red} {\wedge}} \  \, {\textcolor{red} {P^{\geq  .8 } }} \ \,    
  Q _{\leq .4  }     \ \, 
(\vee \ \, R^{\geq .9  }  \ \,   \{\wedge \ \ Q_{\leq .2  }   \  P_{\leq .3  }  \} )  \,   \}$
\qed
\end{example}

   We now extend our analysis from HNCs with  pattern 
 ${\textcolor{red} {X^{\geq   \alpha } }} \, {\textcolor{red}\wedge} \,  
({\textcolor{blue} {\vee}} \ \, {\textcolor{blue} {X_{\leq   \beta }}} \ \, 
 \mathcal{D}({\textcolor{blue} {X_{\leq  \beta  } }}) \,)$ 
to those with   pattern  
 ${\textcolor{red} {X^{\geq   \alpha } }}  \, {\textcolor{red} {\wedge}} \, ({\textcolor{blue} {\vee}} \  \ 
\mathcal{C}({\textcolor{blue} {X_{\leq  \beta } }})  \ \ \mathcal{D}({\textcolor{blue} {X_{\leq   \beta } }})\, )$ 
in which $\mathcal{C}({\textcolor{blue} {X_{\leq    \beta }}})$ is 
\underline{the greatest sub-formula}
\underline{of the input $ \varphi $ that becomes false 
 when ${\textcolor{blue} {X_{\leq    \beta }}}$ is false}, that is, $\mathcal{C}({\textcolor{blue} {X_{\leq    \beta }}})$ is the greatest sub-formula 
 of  $ \varphi $ containing ${\textcolor{blue} {X_{\leq    \beta }}}$ and equivalent to a conjunction  ${\textcolor{blue} {X_{\leq    \beta }}} \wedge \psi$.   
For instance, if the input has the sub-formula
 $\{\wedge \ \phi_3 \ \{\wedge \ {\textcolor{blue} {X_{\leq  \beta  } }} \ 
 (\vee \ \phi_1 \ P^{\geq .3} \, )\} \ \phi_2 \}$, we take it as $\mathcal{C}({\textcolor{blue} {X_{\leq    \beta }}})$ because it is equivalent to 
 $ {\textcolor{blue} {X_{\leq  \beta  } }} \,\wedge \,\{\wedge \ \phi_3 \   \ 
 (\vee \ \phi_1 \ P^{\geq .3} \, ) \ \phi_2 \}$. Thus, if $ {\textcolor{blue} {X_{\leq    \beta }}} $
 becomes false so does the formula $ \mathcal{C}({\textcolor{blue} {X_{\leq    \beta }}}) $.

\vspace{.15cm}
\noindent {\bf Remark.} $\mathcal{C}({\textcolor{blue} {X_{\leq  \beta  } }})$ contains 
  ${\textcolor{blue} {X_{\leq   \beta } }}$
while  $\mathcal{D}({\textcolor{blue} {X_{\leq   \beta  }}})$ excludes  it.

\begin{example} \label{ex:simple-conjuction}  The next formula is an extension of  that  
from Example \ref{ex:withoutconj}, where  
 $ {\textcolor{blue} {P_{\leq .5  } }} $ is substituted by the formula $ \{\wedge \ \, {\textcolor{blue} {P_{\leq .5  } }} \ \, (\vee \ \, R^{\geq .7  }   \ \, Q_{\leq  .1 }  ) \, \} $:
$$\bullet \  \varphi=\{{\textcolor{red} {\wedge}} \  \, {\textcolor{red} {P^{\geq .8  } }} \ \,    
(\vee   \ \,  Q_{\leq  .4 }  \ \,  \{\wedge \ \, {\textcolor{blue} {P_{\leq .5  } }} \ \, (\vee \ \, R^{\geq .7  }   \ \, Q_{\leq  .1 }  ) \, \}\,) \ \ 
(\vee \ \, R^{\geq  .9 }  \ \, 
 \{\wedge \ \ Q_{\leq .2  }   \  P_{\leq  .3 }  \,\} \,)  \,\}$$

\noindent If we select  ${\textcolor{red} {P^{\geq .8  } }} $ and
$ {\textcolor{blue} {P_{\leq .5  } }} $, then   $\varphi$ has   
one sub-formula with the  ${\textcolor{red} {X^{\geq   \alpha }}} \, {\textcolor{red} {\wedge}} \,  ({\textcolor{blue} {\vee}} \   
\mathcal{C}({\textcolor{blue} {X_{\leq  \beta  } }}) \,  \mathcal{D}({\textcolor{blue} {X_{\leq  \beta  } }})\, )$ pattern 
in  which   
$\mathcal{C}({\textcolor{blue} {P_{\leq .5  } }})=\{\wedge \  
{\textcolor{blue} {P_{\leq .5  } }} \ (\vee \  R^{\geq .7  } \  Q_{\leq .1 } ) \, \}$ and
 $\mathcal{D}({\textcolor{blue} {P_{\leq .5  } }})=  (\vee \ Q_{\leq .4 } )$. \qed
\end{example} 
 
 Thus, the  needed  RUR  
for the extended pattern $ {\textcolor{red} {X^{\geq   \alpha }}} \  {\textcolor{red}\wedge} \ 
({\textcolor{blue} {\vee}} \  \mathcal{C}({\textcolor{blue} {X_{\leq  \beta  } }}) 
\  \mathcal{D}({\textcolor{blue} {X_{\leq  \beta  } }}) \,) $  is: 
\begin{equation} \label{eq:simple-conjunction}
\frac{ {\textcolor{red} {X^{\geq  \alpha  } }} \ \, {\textcolor{red}\wedge} \ \,
({\textcolor{blue} {\vee}} \ \, \mathcal{C}({\textcolor{blue} {X_{\leq   \beta } }}) 
\ \, \mathcal{D}({\textcolor{blue} {X_{\leq  \beta  } }}) \,), \  \alpha >  \beta}
{   \mathcal{D}({\textcolor{blue} {X_{\leq  \beta  } }})  }{\ \mbox{RUR}}
\end{equation}

\begin{example} By applying RUR with literals ${\textcolor{red} {P^{\geq .8  } }} $ and  
$ {\textcolor{blue} {P_{\leq .5  } }} $ to  $ \varphi $ in Example \ref{ex:simple-conjuction} 
and then removing the generated redundant 
$ \vee $-connective, we obtain:
$$ \varphi'= \{{\textcolor{red} {\wedge}} \  \, {\textcolor{red} {P^{\geq  .8 } }} \ \,   Q_{\leq  .4 }   \ \ (\vee \ \, R^{\geq  .9 }  \ \, 
 \{\wedge \ \ Q_{\leq .2  }   \  P_{\leq  .3 }  \,\} \,)  \,\}$$
\end{example}

\subsection{General HNC formulas}  \label{subsec:generalHNF}

\vspace{.1cm}
 We  now consider   arbitrarily nested HNCs 
 to which  the rule   RUR can indeed  be applied, which means     HNCs with  the next 
 pattern\footnote{The notation $ \langle \odot \ \varphi_1 \ldots \varphi_k \rangle $ was introduced 
 in Definition \ref{def:NCformulas}, bottom.}:
 $$\{\wedge \ \, \pi_0 \ \,  {\textcolor{red} {X ^{\geq \alpha  } }}  
 \ \, \pi'_0 \ \langle \odot_1 \ \, \pi_1  \ldots    \langle \odot_k   \,   \ \pi_{k}    
\ \ \ ({\textcolor{blue} {\vee}} \ \,  \mathcal{C}({\textcolor{blue} {X_{\leq   \beta  }}})
  \  \mathcal{D}({\textcolor{blue} {X_{\leq  \beta  } }}) ) 
 \ \ \ \pi'_k \, \rangle   \ldots     \pi'_1 \rangle \ \, \pi_0'' \}$$

\noindent where   the $\pi$'s and  $\pi'$'s are concatenations of HNC formulas,  
for instance, for the nesting level $ j $, we have $\pi_j=\varphi_{j_1} \ldots \varphi_{j_{i-1}}$  and 
$\pi'_j=\varphi_{j_{i+1}} \ldots \varphi_{j_{n_j}}.$ 
 It is not hard to check that    RUR can be generalized
simply as follows:
$$\frac{ {\textcolor{red} {X ^{\geq  \alpha   } }}  
\ {\textcolor{red} {\wedge}} \  \langle\odot_1 \ \pi_1     \ldots  \langle \odot_k  
 {\textcolor{red}  \  \pi_{k} \ \  
 ({\textcolor{blue} {\vee}} \   \mathcal{C}({\textcolor{blue} {X_{\leq  \beta  } }})
  \  \mathcal{D}({\textcolor{blue} {X_{\leq  \beta  } }}) )  
\  \ \pi'_k\,\rangle } \ldots   \pi'_1\rangle, \  \alpha >  \beta }
 {  
 \langle \odot_1 \ \pi_1   \ldots  \langle \odot_k   \ \pi_{k} \ \  
  \mathcal{D}({\textcolor{blue} {X_{\leq  \beta  } }})  \ \ \pi'_k\,\rangle 
  \ldots  \pi'_1 \rangle } 
 {\ \mbox{RUR}}$$

{\bf Remark.} RUR should be read as: if the input  $ \varphi $ has any sub-formula    having the 
 pattern of the right  conjunct of the  numerator then it can be
replaced with the formula in the denominator.   
In practice,  applying RUR amounts to simply remove  
$\mathcal{C}({\textcolor{blue} {X_{\leq  \beta  } }})$.

\vspace{.15cm}
 In order to simplify RUR,  we   denote $\Pi$ the right conjunct of the numerator  and   
also denote $\Pi \cdot ({\textcolor{blue} {\vee}} \  \mathcal{C}({\textcolor{blue} {X_{\leq   \beta  }}}) 
\  \mathcal{D}({\textcolor{blue} {X_{\leq    \beta }}})  \,)$  that 
$({\textcolor{blue} {\vee}} \  \mathcal{C}({\textcolor{blue} {X_{\leq  \beta  } }}) 
\  \mathcal{D}({\textcolor{blue} {X_{\leq   \beta } }})  \,)$ is a sub-formula of $\Pi$.  
Using both notations, the  rule  above can be  compacted  giving rise to    RUR for
arbitrary   HNCs:

$$\tcboxmath{{\frac{{\textcolor{red} {X ^{\geq  \alpha  } }} \ {\textcolor{red} {\wedge}} \ 
\Pi \cdot ({\textcolor{blue} {\vee}} \ 
 \ \mathcal{C}({\textcolor{blue} {X_{\leq  \beta  } }}) \ \ \mathcal{D}({\textcolor{blue} {X_{\leq  \beta  } }})  \,), \  \alpha >  \beta  }
{
\Pi \cdot   \mathcal{D}({\textcolor{blue} {X_{\leq  \beta  } }})     }{\mbox{\,RUR}}}}$$

\vspace{.1cm}
In the next   Examples \ref{ex:formulacompl} 
and \ref{ex:cont}, we   analyze the formula $ \varphi $  from the Introduction section (and also from  Example \ref{ex:introduction}).

\begin{example} \label{ex:formulacompl} Let us consider  $ \varphi $  from the 
Introduction  where $ \varphi'= {\textcolor{red} {P^{\geq .7  } }} $. 

\vspace{.15cm}
$\varphi=\{\wedge \ \,  P_{\leq .8  }  \ \,$ 

\vspace{.15cm} 
\hspace{1.2cm}$ (\vee \ \ P_{\leq  .2 }  \ \ \{\wedge \ \ (\vee \ \ P_{\leq .3  } \ \ Q_{\leq .4  }  \ \ P^{\geq  1.  }) \ \ (\vee \ \ \phi_1 \ \ \{\wedge \ \ \phi_2  \ \ {\textcolor{blue} {P_{\leq .6  } }} \}  )$
%
$ \ \, Q^{\geq .7  }  \} \,) $

\vspace{.15cm}
\hspace{1.2cm}$ \ \  {\textcolor{red} {P^{\geq .7  } }} \}$

\vspace{.15cm}
\noindent Its associated DAG (tree, in this is case) is depicted in {\bf Fig. 1} below:
\begin{center}
\begin{adjustbox}{valign=t}
\begin{tikzpicture}[]
\Tree
[.$\wedge$   [.{$P_{\leq .8  } $}\\ ]
             [.$\vee$  [.{\color{black}$P_{\leq .2  } $}\\ ] 
                       [.$\wedge$  [.$\vee$ [.{$P_{\leq .3  } $}\\ ] [.{$Q_{\leq .4  } $}\\ ] 
                        [.{$P^{\geq 1.  } $}\\ ]   ] 
                                   [.$\vee$ [.$\phi_1$\\ ] 
                                              [.$\wedge$ [.$\phi_2$\\ ] [.{\color{blue}$P_{\leq .6  } $}\\ ] ]
                                              ]
                                   [.{$Q^{\geq .7  } $}\\ ] 
                       ]
              ]
             [.{\color{red} { $P^{\geq .7}$}}\\ ]
        ]
\end{tikzpicture}
\end{adjustbox}

\vspace{-.2cm}
{\footnotesize {\bf Fig.  1.} {\em Tree of  Example \ref{ex:formulacompl}.}}
\end{center}


\vspace{.2cm}
\noindent Selecting the literals  ${\textcolor{red} {P^{\geq  .7 } }}$ and  
${\textcolor{blue} {P_{\leq .6  } }}$,  the formula
$\Pi \cdot (\vee  \ \mathcal{C}({\textcolor{blue} {P_{\leq .6  } }}) \ \mathcal{D}({\textcolor{blue} {P_{\leq  .6 } }}))$ is:
$$ (\vee \ \ P_{\leq .2  }  \ \ \{\wedge \  \   
(\vee   \ \   P_{\leq .3  }  \ \ Q_{\leq .4  } \ \ P^{\geq  1. }  \,) \ \ (\vee \ \ \phi_1 \ \  
 \{\wedge \ \ \phi_2  \ \ {\textcolor{blue} {P_{\leq .6  } }} \}  ) \ \ Q^{\geq  .7 }  \} ) $$

\noindent wherein the sub-formula $ (\vee  \  \mathcal{C}({\textcolor{blue} {P_{\leq  .6 } }})
\  \mathcal{D}({\textcolor{blue} {P_{\leq .6  } }}))$ is 
$(\vee \  \phi_1 \  \{\wedge \ \phi_2  \ {\textcolor{blue} {P_{\leq .6  } }} \}  )$. 
 Applying   RUR to $ \varphi $
leads to the next simpler formula:  
$$ \varphi'=\{\wedge \ \ P_{\leq .8  } \ \ (\vee \ \ P_{\leq .2  }  \ \ \{\wedge \  \   
(\vee   \ \   P_{\leq  .3 }  \  \ Q_{\leq .4  }   \ \ P^{\geq 1.  }  \,) 
\ \ (\vee \  \phi_1) \ \ Q^{\geq .7  }  \} ) 
\ \ {\textcolor{red} {P^{\geq  .7 } }} \}$$
\noindent  whose associated tree is   depicted in  {\bf Fig. 2} below.

\begin{center}
\begin{adjustbox}{valign=t}
\begin{tikzpicture}[]
\Tree
[.$\wedge$   [.{$P_{\leq .8  } $}\\    ]
             [.$\vee$  [.{\color{blue}{$P_{\leq .2  } $}}\\ ] 
                       [.$\wedge$  [.$\vee$ [.{$P_{\leq .3  } $}\\ ] [.{$Q_{\leq  .4 } $}\\ ] 
                                                                     [.{$P^{\geq  1.  }$}\\ ] ] 
                                   [.$\vee$ [.$\phi_1$\\ ] 
                                              ]
                                   [.{$Q^{\geq .7  }$}\\ ] 
                       ]
              ]
             [.{\color{red} {$P^{\geq .7}$}}\\ ]
        ]
\end{tikzpicture}
\end{adjustbox}

\vspace{-.2cm}
{\small {\bf Fig.  2.} {Tree of  Formula  $ \varphi' $ }}
\end{center}
\vspace{-.6cm}
\qed
\end{example}

\noindent  $\bullet$ {\bf Simplification Rules.} To complete the calculus  $\mathit{RUR_{NC}}$,  
the simplification
 rules given below must  accompany {\em RUR. }
 Recall that, by Definition \ref{def:constant},  
 $ (\vee) $ is a ${\bot}$-constant.
  The first two rules  
  simplify  formulas by (upwards) propagating  $(\vee)$   from sub-formulas to formulas,
in which  $ \varphi $ is the input HNC, and, as before,  $ \varphi \cdot \phi $ means that  $ \phi $ is a sub-formula
  of $ \varphi $:

\begin{equation} \label{inf:vee-vee}
\frac{  \varphi \cdot ({\textcolor{black} {\vee}} \  \ \phi_1 \ldots   \phi_{i-1} 
\, (\vee) \, \phi_{i+1} \ldots \phi_k  \,)  }
{ \varphi \cdot ({\textcolor{black} {\vee}} \  \ \phi_1 \ldots   \phi_{i-1} 
\,  \phi_{i+1} \ldots \phi_k  \,)  }{\vee \bot}
\end{equation}

\begin{equation} \label{inf:wedge-wedge}
\frac{
\varphi \cdot \{{\textcolor{black} {\wedge}} \  \ \varphi_1 \ldots   \varphi_{i-1} 
\, (\vee) \, \varphi_{i+1} \ldots \varphi_k  \,\}  }
{ \varphi \cdot  (\, {\textcolor{black} {\vee}} \,)  }{\wedge \bot}
\end{equation}

\noindent The next two rules  remove redundant connectives; the first one 
removes a connective  if it is applied to a
single sub-formula,
i.e. $ \langle \odot_1 \  \phi_1 \,\rangle $:

\begin{equation} \label{inf:odot}
\frac{ \varphi \cdot \langle{\textcolor{black} {\odot_2}} \  \ \varphi_1 \ldots   \varphi_{i-1} 
\, \langle \odot_1 \  \phi_1 \,\rangle \, \varphi_{i+1} \ldots \varphi_k  \,\rangle  }
{\varphi \cdot \langle{\textcolor{black} {\odot_2}} \  \ \varphi_1 \ldots   \varphi_{i-1} 
 \,  \phi_1 \,  \varphi_{i+1} \ldots \varphi_k  \,\rangle }{ \odot_{k+1}}
\end{equation}

\noindent and the next rule removes a connective  of a sub-formula 
  that is inside 
another sub-formula with the same  connective: 
\begin{equation} \label{inf:odot}
\frac{ \varphi \cdot \langle{\textcolor{black} {\odot_1}} \  \ \varphi_1 \ldots   \varphi_{i-1} 
\, \langle\odot_2 \  \phi_1 \ldots \phi_n \,\rangle \, \varphi_{i+1} \ldots \varphi_k  \,\rangle, \odot_1=\odot_2  }
{\varphi \cdot \langle{\textcolor{black} {\odot_1}} \  \ \varphi_1 \ldots   \varphi_{i-1} 
 \,  \phi_1 \ldots \phi_n \,  \varphi_{i+1} \ldots \varphi_k  \,\rangle }{\odot_{k + n}}
\end{equation}

\noindent Finally,  two further trivial simplification  rules are  required: 
$$\frac{\{\wedge \  P^{\geq  \alpha  } \  P_{\leq   \beta }  \,\},  \alpha >  \beta}
{{(\vee)}}{\bot ^\alpha_\beta} 
 \quad \ \ 
 \frac{\{\wedge \  P^{\geq  \alpha  }  \  P^{\geq   \beta }  \,\}, \gamma=\mathrm{max}( \alpha, \beta)}{P^{\geq \gamma  } }{\,\mathrm{max}}$$

\noindent {\bf Remark}. Rules like the following ones: 
$$\frac{(\vee \  P^{\geq  \alpha  }  \  P_{\leq  \beta  }  \,),  \alpha \geq  \beta   }{{\top}} \quad \ \ \ \frac{(\vee \  \phi_1 \ldots \phi_k \ P_{\leq   \alpha }  \  P_{\leq  \beta  }  ), 
\gamma=min( \alpha, \beta)}
{(\vee \  \phi_1 \ldots \phi_k \ P_{\leq \gamma  } )} $$

\noindent are of course sound and  have indeed interest for improving efficiency but 
are unnecessary 
for warranting refutational completeness,
which is our concern in this paper.

\begin{definition} We  define    $\mathit{RUR_{NC}}$ as the  calculus
 formed by  the  rule RUR and the above described simplification rules, namely,  
 $$\mathit{RUR_{NC}}= \{\mathit{RUR},\, \vee\bot, \, \wedge\bot, \,  \odot_{k+1}, \,  \odot_{k+ n},\,   \bot^\alpha_\beta, \,\mathrm{max} \}$$
\end{definition}

\begin{example} \label{ex:cont} By continuing   Example  \ref{ex:formulacompl},
we  now show  how   $\mathit{RUR_{NC}}$ finds that $\varphi'  $ is  unsatisfiable.
 If we select ${\textcolor{red} {P^{\geq  .7 } }}$ and   ${\color{blue}{P_{\leq .2  } }}$ 
(colored blue  in {\bf Fig. 2}),   then:  
$\mathcal{C}({\color{blue}{P_{\leq .2  } }})={\color{blue}{P_{\leq .2  } }}$;   
\  $\mathcal{D}({\color{blue}{P_{\leq .2  } }})=(\vee \ \{\wedge \  \,   
(\vee   \ \,   P_{\leq .3  }  \  \, Q_{\leq  .4  } \ \, P^{\geq 1} \,) 
\  (\vee \ \, \phi_1) \ \, Q^{\geq .7  }  \} \,)$; and
$$\Pi \cdot (\vee \ \mathcal{C}({\color{blue}{P_{\leq .2  } }}) 
\ \mathcal{D}({\color{blue}{P_{\leq  .2 } }})) =
(\vee \ \mathcal{C}({\color{blue}{P_{\leq .2  } }}) \ \mathcal{D}({\color{blue}{P_{\leq .2  } }}))$$  

\noindent By applying   RUR to $ \varphi' $,  
 the obtained formula $ \varphi'' $ is depicted  in    {\bf Fig. 3} below: 
\begin{center}
\begin{adjustbox}{valign=t}
\begin{tikzpicture}[]
\Tree
[.$\wedge$   [.{$P_{\leq .8  } $}\\   ]
             [.$\vee$   
                       [.$\wedge$  [.$\vee$ [.{$P_{\leq .3  } $}\\ ] [.{$Q_{\leq .4  } $}\\ ] [.{$P^{\geq  1. } $}\\ ] ] 
                                   [.$\vee$ [.$\phi_1$\\ ] 
                                              ]
                                   [.{$Q^{\geq  .7 } $}\\ ] 
                       ]
              ]
             [.{\color{red} { $P^{\geq .7}$ }}\\ ]
        ]
\end{tikzpicture}
\end{adjustbox}

\vspace{-.15cm}
{\small {\bf Fig.  3.} {Tree of  Formula  $ \varphi'' $ }}
\end{center}

\noindent After two applications of  $\odot_{k+ 1}$ and one of $\odot_{k+ n}$, one gets the formula 
associated with the   tree in  {\bf Fig. 4} below:

\begin{center} 
\begin{adjustbox}{valign=t} \hspace{2.cm}
\begin{tikzpicture}[]
\Tree
[.$\wedge$   
                         [.{$P_{\leq .8  }  $}\\    ]
                         [.$\vee$ [.{\color{blue} {$P_{\leq  .3  }$}}\\ ] 
                         [.{\color{blue} {$Q_{\leq  .4 } $}}\\ ] [.{$P^{\geq 1.  } $}\\ ] ] 
                                    [.$\phi_1$\\ ] 
                                   [.{\color{red} {$Q^{\geq  .7 } $}}\\ ] 
             [.{\color{red} { $P^{\geq .7}$}}\\ ]
        ]
\end{tikzpicture}
\end{adjustbox}

{\small {\bf Fig. 4.} {\em Continuing Example \ref{ex:cont}}} 
\end{center}

\noindent
In this state, two  applications of  RUR  to the two complementary  pairs {\color{red} {$Q^{\geq .7} $}} 
and {\color{blue} {$Q_{\leq .4} $}}, and   
{\color{red} {$P^{\geq .7}$}} and {\color{blue} {$P_{\leq .3} $}},  
and then one application of $\odot_{k+ 1}$
lead the calculus $\mathit{RUR_{NC}}$ to  the formula associated with the tree despited in {\bf Fig. 5}, left.

\begin{center} 
\begin{adjustbox}{valign=t} 
\begin{tikzpicture}[]
\Tree
[.$\wedge$   
                         [.{$P_{\leq  .8 }  $}\\    ]
                            [.{\color{red} {$P^{\geq 1.  } $}}\\ ]  
                                    [.$\phi_1$\\ ] 
                                   [.{\color{black} {$Q^{\geq .7  }$}}\\ ] 
             [.{\color{red} {$P ^{\geq .7}$}}\\ ]
        ]
\end{tikzpicture}
\end{adjustbox}
\begin{adjustbox}{valign=t} \hspace{.3cm}
\begin{tikzpicture}[]
\Tree
[.$\wedge$   
                         [.{\color{blue} {$P_{\leq  .8 }  $}}\\    ]
                            [.{\color{red} {$P^{\geq  1. } $}}\\ ]  
                                    [.$\phi_1$\\ ] 
                                   [.{\color{black} {$Q^{\geq .7  } $}}\\ ] 
        ]
\end{tikzpicture}
\end{adjustbox}

{\small {\bf Fig. 5.} {\em Continuing Example \ref{ex:cont}}} 
\end{center}

\noindent Now, the inference rule, called  max, is applied and the 
formula inferred has the right tree in {\bf Fig. 5}. Finally, the rule $\bot^\alpha_\beta  $ derives  
  $(\vee)$. \qed
\end{example}

\begin{lemma} An HNC $\varphi$ is unsatisfiable iff     $\mathit{RUR_{NC}}$ applied  to $\varphi$ derives 
$ (\vee )$.
\end{lemma}

\begin{niceproof} We analyze below both senses of the lemma.

\vspace{.1cm}
$\bullet \ \Rightarrow $ Let us assume that $\varphi$ is unsatisfiable. Then $\varphi$ must have
a sub-formula verifying the RUR numerator, otherwise, all unsatisfiable pairs of literals 
$X^{\geq  \alpha  }$ and  $X_{\leq  \beta  }$ such that $ \alpha >  \beta$  are 
included in disjunctions (if $ \alpha =  \beta$, both literals are satisfiable). If this case, since all disjunctions of $ \varphi $, 
by definition of HNC formula, have
at least one negative literal, $ \varphi $ would be satisfied by assigning to all propositions
 the value 0,
which contradicts the initial hypothesis. Therefore, RUR is applied to $ \varphi $ with 
the literals $X^{\geq  \alpha  }$ and  $X_{\leq  \beta  }$ such that $ \alpha >  \beta$ 
and the resulting formula is simplified. The new  formula is equivalent to $ \varphi $ 
and has at least one literal less than $ \varphi $. Hence, by induction on the number of literals
of $ \varphi $, we easily obtain that $\mathit{RUR_{NC}}$  ends only when $ (\vee )$ is derived.

 $ \bullet \Leftarrow$   Let us assume that  $\mathit{RUR_{NC}}$ has been applied 
without having derived  $ (\vee )$. Clearly, 
   $\mathit{RUR_{NC}}$ ends when  a deduced formula $\varphi'$  
is not a conjunction of a literal $X^{\geq  \alpha  }$ with  a disjunction 
 including another literal $X_{\leq  \beta  }$ such that $ \alpha >  \beta$.
 Firstly, since $\mathit{RUR_{NC}}$ is sound, $ \varphi $ and   $ \varphi' $ are equi-satisfiable.
Secondly, if $\varphi'$ has   complementary literals, then they are integrated in
 disjunctions. Thus  
 $\varphi'$ is satisfied by assigning the value 0 to all its unassigned propositions, since, 
 by definition of HNC formula, 
 all  disjunctions 
have at least one negative  disjunct. 
 Therefore, since $\varphi'$ is satisfiable so is  $\varphi$.
\end{niceproof}

\begin{lemma} \label{lem:polytime} Reg-Horn-NC-SAT is polynomial.
\end{lemma}

\begin{niceproof} The number of inferences performed by $\mathit{RUR_{NC}}$ 
to the input  $ \varphi $ is
bounded linearly in the size of $ \varphi $. Indeed, the number of   RUR rules performed 
is at most the     number of literals
in $ \varphi $, and the number of  simplification rules  is  at most the   number 
of connectives plus the number of literals in $ \varphi $.
On the other hand, it
is not difficult to find data structures to polynomially execute  each inference
of the calculus $\mathit{RUR_{NC}}$. Hence, Reg-Horn-NC-SAT is polynomial.
\end{niceproof}

\begin{proposition} Reg-Horn-NC-SAT is P-complete.
\end{proposition}

\begin{niceproof} It follows straightforwardly from the next two facts: 
(i) Reg-Horn-NC-SAT is polynomial, according to Lemma
\ref{lem:polytime}; and (ii) Reg-Horn-NC-SAT includes   Reg-Horn-SAT which in turn
includes  propositional Horn-SAT which is  P-complete \cite{BryEEFGLLPW07}.
\end{niceproof}

\noindent {\bf Remark.} Having established $\mathit{RUR_{NC}}$, the procedure NC Unit-Propagation  
(viz. the  Reg-Horn-NC-SAT algorithm) can 
easily  be designed, and on its basis,   effective  Non-Clausal DPLL-based solvers can be developed.

\vspace{.2cm}
 The best published complexities  for    Reg-Horn-SAT are $ O(n  \log n) $  and $ O(n)$
for infinite- and finite-valued regular formulas, respectively   \cite{Hahnle96,BeckertHahnleManya00,Hahnle01,Hahnle03}.  
For future work,  we   will  attempt to find
  data structures and devise   algorithms, inspired by 
   $\mathit{RUR_{NC}}$, able  to decide  
 the satisfiability of HNCs with   complexity  
  tight  to the aforementioned clausal ones.

\subsection{Further Inferences Rules}  \label{subsec:Further-infer}

In this subsection, we present  two further inferences extending $\mathit{RUR_{NC}}$:
Regular General-Unit-Resolution and Regular Hyper-Unit-Resolution.

\vspace{.15cm}
\noindent {\bf General Unit-Resolution.} So far we have defined   $\mathit{RUR_{NC}}$        just for  HNCs and thus,
 its design    for general NCs was pending. For such purpose, 
 the application of the RUR rule to  sub-formulas $ \phi $ 
of the input NC $ \varphi $ having the RUR numerator pattern should be authorized. Namely,
applying the previous RUR rule to each sub-formula  of  $\varphi $  with pattern   
 $\phi={\textcolor{red} {X^{\geq  \alpha}}} \ {\textcolor{red} {\wedge}} \ 
\Pi \cdot ({\textcolor{blue} {\vee}} \  
 \mathcal{C}({\textcolor{blue} {X_{\leq  \beta}}}) \ 
  \mathcal{D}({\textcolor{blue} {X_{\leq  \beta}}})  \,)$   should be permitted
 and so, $ \phi $ could be replaced  with 
${\textcolor{red} {X^{\geq  \alpha}}} \ {\textcolor{red} {\wedge}} \ 
\Pi \cdot  
  \mathcal{D}({\textcolor{blue} {X_{\leq  \beta}}}) $.  
Hence,     the formal specification of the  Regular General-Unit-Resolution 
rule, RGUR, for any general NC   $ \varphi $ is:

$$\tcboxmath{\frac{\varphi \cdot (\,{\textcolor{red} {X^{\geq  \alpha}}} \ {\textcolor{red} {\wedge}} \ 
\Pi \cdot ({\textcolor{blue} {\vee}} \  
 \mathcal{C}({\textcolor{blue} {X_{\leq  \beta}}}) \ 
  \mathcal{D}({\textcolor{blue} {X_{\leq  \beta}}})  \,) \,) }
{\varphi \cdot (\,{\textcolor{red} {X^{\geq  \alpha}}} \ {\textcolor{red} {\wedge}} \ 
\Pi \cdot    \mathcal{D}({\textcolor{blue} {X_{\leq  \beta}}})  \,)  }{\mbox{\,RGUR}}}$$

\begin{example} Consider again  Example \ref{ex:cont}.
 One can check that   its sub-formula
$$ (\vee \ \ P_{\leq  .2 }  \ \ \{\wedge \ \ (\vee \ \ P_{\leq .3  } \ \ Q_{\leq .4  }  \ \ P^{\geq  1.  }) \ \ (\vee \ \ \phi_1 \ \ \{\wedge \ \ \phi_2  \ \ {\textcolor{blue} {P_{\leq .6  } }} \}  )$$ has the  pattern of the RGUR numerator, and
so  RGUR can be applied   and replaced  it with
$ (\vee \ \ P_{\leq  .2 }  \ \ \{\wedge \ \ (\vee \ \ P_{\leq .3  } \  P^{\geq  1.  }) \ \ (\vee \ \ \phi_1 \ \ \{\wedge \ \ \phi_2  \ \ {\textcolor{blue} {P_{\leq .6  } }} \}  )$ in the  input formula. \qed
\end{example}

 Observe that the introduction of this new RGUR rule applicable to certain
sub-formulas habilitates new sequences of inferences, and so, their suitable management  
can enhance the overall deductive efficiency.

\begin{proposition} Let  $ \varphi $ be any NC formula. If $ \varphi' $ 
results from  applying RGUR to $ \varphi $ then $ \varphi' $  and  $ \varphi $ are logically equivalent.
\end{proposition}

\begin{niceproof} The proof of the  soundness of RGUR is straightforward.
\end{niceproof}

\noindent {\bf Remark.} The  extension of the simplification rules 
 from HNCs to NCs is similarly obtained and 
so is the  calculus $\mathit{RGUR_{NC}}$.

\vspace{.2cm}
\noindent {\bf Hyper Unit-Resolution.}  
The given definition of Regular NC  Unit-Resolution  can be  extended in order to
obtain Regular NC Hyper-Unit-Resolution ($\mathit{RHUR}_{NC}$),
which is given below.  The sub-formula
 $({\textcolor{blue} {\vee}} \ 
 \ \mathcal{C}({\textcolor{blue} {X_{\leq   \beta  }}}) 
 \ \ \mathcal{D}({\textcolor{blue} {X_{\leq   \beta  }}})  \,) $ 
 is denoted $ \mathcal{CD}({\textcolor{blue} {X_{\leq   \beta  }}}) $  and since
 $ {\textcolor{blue} {X_{\leq  \beta_i  } }}, 1 \leq i\leq k $, are 
 literal occurrences that are pairwise different, so are  
$ \Pi_i \cdot 
\mathcal{CD}({\textcolor{blue} {X_{\leq  \beta_i  } }}), 1 \leq i\leq k $:

$$\tcboxmath{\frac{{\textcolor{red} {X ^{\geq  \alpha  } }} \ {\textcolor{red} {\wedge}} \ \Pi_1 \cdot 
\mathcal{CD}({\textcolor{blue} {X_{\leq  \beta_1  }}}) \, {\textcolor{red} {\wedge}} \ldots 
\Pi_i \cdot 
\mathcal{CD}({\textcolor{blue} {X_{\leq  \beta_i  } }}) \ldots {\textcolor{red} {\wedge}} \, \Pi_k \cdot 
\mathcal{CD}({\textcolor{blue} {X_{\leq   \beta_k }}}), \forall i,  \alpha >  \beta_i  }
{ \Pi_1 \cdot 
\mathcal{D}({\textcolor{blue} {X_{\leq  \beta_1  } }}) \, {\textcolor{red} {\wedge}} \ldots 
\Pi_i \cdot 
\mathcal{D}({\textcolor{blue} {X_{\leq  \beta_i  } }}) \ldots {\textcolor{red} {\wedge}} \, \Pi_k \cdot 
\mathcal{D}({\textcolor{blue} {X_{\leq   \beta_k } }})    } {RHUR_{NC}}}$$


The soundness and completeness of $\mathit{RHUR}_{NC}$ follow  from
those of $\mathit{RUR}_{NC}$.


\section{Non-Clausal Logic-Programming} \label{sec:logicprogram}


In spite of both the  profusion  of  many-valued logic programming approaches 
in  clausal form  developed since annotated logic programming   was conceived \cite{KiferS92}
  and   the important  advances  carried out in propositional
logic programming  in NC form 
since nested logic programing was proposed
 \cite{LifschitzTT99},
  no approach  seems to have been developed to deal with  many-valued logic programming in NC form. 
 Regarding computational issues,
  tractability is analized, to our knowledge, only in  \cite{imaz2021horn}
  which focuses on propositional logic.
 
 \vspace{.05cm}
 This section  shows the usefulness of both $\mathcal{RH}$ and  
$\mathit{RUR_{NC}}$ for  NC  logic-programming in   regular-logic. Concretely, we show that
the rule syntax  can be enriched allowing  NCs in heads and bodies with slight restrictions while keeping    
query-answering efficiency qualitatively  comparable to   clausal efficiency, 
specifically  answering queries  
 is polynomial.

\begin{definition} \label{def:NCrules}  An HNC  rule is an expression 
 $  B^+ \rightarrow H $ wherein $  B^+ $ (Body) is  
an NC formula having only positive literals
and $H$ (Head) is any arbitrary HNC formula. An HNC logic program is a set of HNC rules.
\end{definition}

\begin{example} \label{ex:NCrule} The next rule is an HNC rule, where its body 
is an NC  with only positive literals and its head is a simple HNC formula.
$$\bullet \ \  \{\wedge \ R^{\geq 1}  \ (\vee \ P^{\geq .7}  \ \{\wedge \ S^{\geq  .7} 
 \ Q^{\geq  .6} \} ) \} \ \longrightarrow \ (\vee  \     \{\wedge \   Q_{\leq .6  }  \   S_{\leq .7  } \} \   
\{\wedge \   R^{\geq  .7 }  \    P^{\geq .3  }  \,\} \,)$$

\noindent Another  HNC rule is given below, where its  body is again a positive NC  
and its head $ \varphi $ can be,   for instance, any HNC  among the ones used  
 in previous illustrative examples:
 
 \vspace{.2cm}
\hspace{0.45cm}$\bullet \ \  \{\wedge \ R^{\geq } 1 \ (\vee \ P^{\geq } .7 \ \{\wedge \ S^{\geq  .7} 
 \ Q^{\geq  .6} \} )  \       
(\vee \ \, R^{\geq  .9 }  \ \,   \{\wedge \ \ Q^{\geq .2  }   \  P^{\geq  .3  } \} ) \ \, Q^{\geq .7  }  \} \longrightarrow \varphi$  \qed

\end{example}

\begin{proposition} A conjunction  of  HNC  rules, or equivalently, 
an HNC logic program, is an HNC formula.
\end{proposition}

\begin{niceproof} Clearly a rule $ (\vee \  \neg B^+  \ H )$ verifies Definition \ref{theorem:visual} and hence 
 so does a conjunction thereof, or equivalently, so does an HNC logic program.
\end{niceproof}

The  rules from Example \ref{ex:NCrule} give an intuitive idea of  the potentiality of $\mathcal{RH}$ 
to  enrich  declarative rules. 
Next lemma analyzes the dual aspect to expressiveness, i.e.  efficiency, 
 stating that  the complexity of query-answering is polynomial (as in the clausal framework).  

\begin{lemma} \label{lemm:Complexity-Program}   Let $ \mathcal{S} $ be a  positive literal set,  
 Lp  be an HNC logic  program and $ \varphi $ be any  arbitrary (unrestricted) NC formula. 
 Deciding whether $\mathcal{S}  \wedge Lp \models \varphi $   is polynomial.
\end{lemma}

\begin{niceproof}  By Lemma  \ref{lem:polytime},  one  can polynomially check  
whether $ \mathcal{S}  \wedge Lp$ is satisfiable. If  so, then by applying $\mathit{RUR_{NC}}$,
 one can polynomially obtain the positive 
literals that follow logically from 
$ \mathcal{S}  \wedge Lp$, i.e. its minimal model.  
Finally, one can 
 also polynomially check    whether,
  for such minimal model, $ \varphi $  is evaluated to  1, i.e. whether $\mathcal{S}  \wedge Lp \models \varphi $  holds.
\end{niceproof}

Summarizing, the  advantages conferred by $\mathcal{RH}$  and $\mathit{RUR_{NC}}$
   to NC logic programs are that,
clearly NC logic programs are smaller and can be even exponentially smaller than their equivalent clausal  logic programs, and that, according to Lemma \ref{lemm:Complexity-Program}, 
 query-answering  takes polynomial time. 

 \section{Related Work}  \label{sec:relatedwork}

 This research work heavily relies on  the areas of 
  regular   logic, NC reasoning and Horn formulas. In \cite{imaz2021horn},
    related work to NC reasoning and Horn formulas  is 
    extensively discussed, and in this section, we will only discuss related work to regular logic.
  
 Since regular logic  is a (relevant)  sub-class of signed logic, 
 we next start by   introducing signed logic and its variants, and by discussing general aspects. Subsequently we review the existing complexity sub-classes in signed clausal satisfiability
and then  the published methods  for solving signed non-clausal satisfiability.
  
\subsection{General Presentation}

We recall (Section \ref{sec:NCbasis}) that  
signed logic differs from propositional logic only at the literal 
  level. 
  A signed literal
  is a pair $ S \cdot P $ where $  P \in \mathcal{P}$  and $S  $
  is a (usually finite) subset $  S \subseteq \mathcal{T} $ 
   and is satisfied by  $ \mathcal{I} $ only if
  $ \mathcal{I} (P) \in  S$. The remaining concepts 
  given in Sections \ref{sec:NCbasis} and  \ref{sec:Non-clausal} are equally applicable to signed
   logic. $ S $  is called  sign of  $ S \cdot P $ and the idea of using truth-value sets as signs is due to independently  several
 authors \cite{Hahnle90, Doherty91,MurrayR91}. 
  
  \vspace{.05cm}
   Regular logic is  the most studied and employed sub-class of signed logic and
   derives from it when  $ \mathcal{T} $ is  totally ordered and   
  the signs  are  only  of two kinds: 
  $ [- \infty \,, \alpha ]$, which corresponds to $P_{\leq \alpha}  $, or $ [ \alpha \,, \infty ]$,
  which corresponds to $ P^{\geq \alpha} $.
Other main signed sub-classes are  issued  when
    $ \mathcal{T} $ is partially ordered, specially when $ \mathcal{T} $ is a lattice, 
  and when the signs are singletons;  the latter is   called mono-signed logic. 
 Subsection \ref{subsec:relatedW:Clausal}
   reviews the  published  polynomial and NP-complete  
  fragments of signed clausal logic.

    \vspace{.05cm}
 An outstanding feature  \cite{Hahnle94a} of signed clausal forms is that they offer a suitable logical framework for automated reasoning 
  in multiple-valued logics given that  the satisfiability problem of any finitely-valued propositional logic, as well as of certain infinitely-valued logics, is
polynomially reducible to the satisfiability problem of signed clausal formulas. 
On the other hand, the  significance of 
  regular logic comes from the next property demonstrated by Hahnle  \cite{Hahnle96}:
{\em  every signed formula is logically equivalent to some regular formula.} 
 
    \vspace{.05cm} 
 Concerning  applications, signed logic and annotated logic programs \cite{KiferS92} 
   are closely connected to each other \cite{Lu96}. 
   Annotated logic programming is obtained from signed logic when 
   the formulas are Horn clausal and  $ \mathcal{T} $ is a lattice. Annotated logic programming
  is  a suitable language   to manage locally inconsistent,  
   incomplete and uncertain databases, and indeed a large number of systems
    have been developed during the  last  decades.
   
   \vspace{.05cm} 
   Another main application of signed logic is found interpreting a literal  $ S \cdot P $
    as "$ P $ is constrained to the values in $S$".  This allows 
signed clausal forms to be used as a powerful knowledge representation language for 
constraint programming \cite{LuCalmetSch97} and they have shown to be a practical and competitive approach to solving combinatorial decision problems 
\cite{FrischPDN05,BejarMCFG07,AnsoteguiLLM07}.

As    aforementioned,   a great deal of research on signed  logic  
 has been conducted    in  satisfiability
  solving,   logic programming and constraint solving,
  but       signed logic and its variants have also  been
  handled, during the last twenty years, in 
    many  aproximate reasoning  scenarios  such as model-based diagnosis \cite{feldman06multi},
  signed optimization \cite{AnsoteguiBLM13},  signed randomiation \cite{BallersteinT14}, 
    combining signed logic and  linear integer arithmetic
  \cite{AnsoteguiBMV15}, learning  in CSP \cite{VekslerS16}, 
 comparing  resolution proofs and CDCL-with-restarts \cite{Mitchell17}, 
 regular belief merging \cite{Dellunde20},   
   in the   formalization of a recent   real-world multivalued-logic 
  \cite{Fagin20} and more.

\subsection{Signed Clausal} \label{subsec:relatedW:Clausal}

Outside signed logic, the complexity of fuzzy logics  is NP-complete 
or harder \cite{Hanikova11} and  only few and 
very restricted classes are polynomial 
\cite{BofillManyaVidalVillaret15b,BofillManyaVidalVillaret19}.  
Next  we discuss  computational issues in signed  logic and its variants
and show that a variety of polynomial classes exist.

\vspace{.05cm}
The central role  of signed clausal formulas  in automated 
deduction  pointed out above, justified a detailed study of its sub-classes, including algorithms for
and complexities of associated satisfiability problems.  These results are summarized next, and for more 
details, the interested reader may consult  
\cite{BeckertHahnleManya00,Hahnle01, Hahnle01a, Hahnle03} and the references therein.

\vspace{.05cm}
{\bf Signed Resolution.} Efficient decision procedures for signed logic are mostly based on  
the extension of resolution to  signed logic, which appeared  
  in \cite{Hahnle93,Hahnle94a} and independently in \cite{MurrayR91,MurrayRosenthal93A}, 
  also see \cite{BaazFS01}.
  Regular resolution and regular unit-resolution  were given in \cite{Hahnle96}.  Resolution
  when $ \mathcal{T} $ is only partially ordered was proposed by several authors
  \cite{LuMurrayRosenthal98,GanzingerS00,Sofronie-Stokkermans07}. 
  The approach in \cite{GanzingerS00,Sofronie-Stokkermans07} consists in encoding  
  regular formulas
  in   first-order theories with transitive relations and 
  then applying ordered resolution \cite{BachmairG98}
  which  includes transitivity axioms. Mono-signed resolution was developed
   in \cite{BaazF95,BaazFS01} and  in  \cite{GanzingerS00,Sofronie-Stokkermans07}; 
  in the latter approach the authors
  translate a formula into first-order equations and then use superposition calculus \cite{BachmairG94}.

\vspace{.05cm}
{\bf Complexity Classes.} Next we  overview the complexity of sub-classes 
of signed clausal logic. We denote Signed-SAT the satisfiability problem of signed formulas, and Signed-2-SAT denotes Signed-SAT
restricted to formulas whose  clauses have only two literals.  
Reg-SAT (Section \ref{sec:NCbasis}) and Reg-2-SAT   are similarly
defined for regular logic. 

\vspace{.1cm}
{\bf  NP-Completeness.} Signed-SAT is NP-complete: its NP-membership is verified as for classical SAT and 
 its  NP-completeness is obvious because it includes classical SAT. Both  
Signed-2-SAT  and Reg-2-SAT for any $ \vert \mathcal{T} \vert \geq 3$ are  
NP-complete  \cite{BeckertHahnleManya99, Manya00}, 
in contrast with the linearity of classical 2-SAT \cite{AspvallPlassTarjan79}.
 Even Reg-2-SAT is NP-complete 
for general partial orders, which is proved by reducing Signed-2-SAT to Reg-2-SAT \cite{BeckertHM00a}.
More concretely, in \cite{BeckertHM00a} it is proved that Signed-2-SAT is NP-complete
(1) if $ \mathcal{T} $ is a modular lattice and the signs $ S $ are complements
of regular signs $ \geq \alpha $  and $ \leq \alpha $  of $ \mathcal{T} $
(i.e. the literals are $ P_{\leq \alpha} $ and $ P^{\geq \alpha} $), or (2) if 
$ \mathcal{T} $ is a distributive lattice and the signs are regular signs of $ \mathcal{T} $
and their complements. 
Further complexities   on Signed-2-SAT 
based on the Helly property are obtained in \cite{ChepoiCreignouHermannSalzer08,ChepoiCHS10}
and are  discussed below.

\vspace{.05cm}
{\bf Polynomial Signed-2-SAT.} Under certain restrictions, however, Reg-2-SAT is polynomial, 
e.g. when $ \mathcal{T} $
is  totally ordered,   Reg-2-SAT is $ O(n \log n) $, $ n $ being the length of $ \varphi $, which is proved via a reduction to classical 2-SAT \cite{BejarHM01}. A polynomial result for the more general case,
  when $ \mathcal{T} $ is a lattice and all occurring signs are of the form 
  $ \leq \alpha $ or $ \geq \alpha $ 
   ($ P_{\leq \alpha} $ or $ P^{\geq \alpha} $) is in \cite{BeckertHM00a}.  
    Charatonik and Wrona \cite{CharatonikW07} showed that this problem can be solved 
    in quadratic time in the size of the input and in linear 
    time in the size of the formula if the lattice is fixed.  
    For this,  they used  a reduction of a many-valued satisfiability problem on a lattice
    to   classical SAT.
    Somewhat different multi-valued 2-SAT problems are studied  in \cite{CharatonikW07}. 
    In \cite{BaazF95}, the authors proved  that mono-signed 2-SAT is polynomial
    and in \cite{Manya00} a   linear-time procedure for such problem is described. 
    In \cite{AnsoteguiManya03}, regular and mono-signed logics are
merged and  new 2-SAT problems are defined proving   some
of them  are polynomial.

  \vspace{.05cm}
{\bf Polynomial Reg-Horn-SAT.} In the regular Horn-SAT  problem,   three 
have been analyzed  depending on the structure of $ \mathcal{T} $: (1) 
$ \mathcal{T} $ is totally ordered, (2) $ \mathcal{T} $ is a lattice, and (3) $ \mathcal{T} $
is partially ordered but   not a lattice.

(1)  When $ \mathcal{T} $ is totally ordered, Reg-Horn-SAT can be solved in
  time linear in $size (\varphi) $ if  $ \mathcal{T} $ is finite, and in $ O(n \log n )$
  otherwise \cite{Hahnle96,BejarHM01}. 
  Many of the results are proven via reduction to classical logic \cite{BejarHM01}.
  Some Horn problems are defined in  \cite{AnsoteguiManya03}
  for the aforementioned merged regular and mono-signed logic 
   and their tractability are stated.
  Complementary results are obtained in \cite{GilHSZ04,GilHSZ08}, 
  where, given a set of interpretations $ M $, the authors 
  obtain a regular Horn 
  formula  whose set of models is $ M $ (called constraint description problem).
  
 (2) If  $ \mathcal{T} $ is a finite lattice, then  Reg-Horn-SAT is decidable in linear time 
  in the size of the formula and polynomial in the cardinality of $ \mathcal{T} $ via 
  a reduction to classical Horn-SAT \cite{BeckertHahnleManya99}. For distributive lattices,
  the bound obtained in \cite{Sofronie-Stokkermans98} is $ size(\varphi) \times n^2 $, where $ n $ is the cardinality
  of $ \mathcal{T} $. A closer inspection of the proofs in the cited paper yields immediately that
  all  Reg-Horn-SAT  problems with fixed size truth values  have linear 
  complexity. \vspace{.05cm}
If $ \mathcal{T} $ is infinite, then Reg-Horn-SAT is decidable provided that 
$ \mathcal{T} $ is a locally finite lattice, that is, every sub-lattice 
generated by a finite subset is finite \cite{BeckertHahnleManya99}.

  \vspace{.05cm}
 (3) If the partial order of $ \mathcal{T} $ is not a lattice,   then a natural notion
  of regular sign can still be obtained by using signs of the form 
  $ S^\geq =\{i \in  \mathcal{T} \mid \exists j \in S \ \mbox{s.t.} \ i \geq j\}$, where 
  $ S \subseteq \mathcal{T} $. This more general  Reg-Horn-SAT is still decidable
   linearly in the length of the formula, but exponential in the cardinality of $ \mathcal{T} $
  provided that $ \mathcal{T} $ possesses a maximal element \cite{BeckertHahnleManya99}.

\vspace{.05cm}
{\bf Helly Property.} We  denote $ \mathcal{S} $ the set of signs 
occurring in a given formula $ \varphi $.
   While in the previous works order-theoretic properties of the truth-value domain $ \mathcal{T} $
    are exploited
to make conclusions on the complexity of signed SAT problems,
Chepoi et al. \cite{ChepoiCreignouHermannSalzer08,ChepoiCHS10} completely settle the complexity question in the general case 
by reverting to combinatorial properties of the set system $ \mathcal{S} $. In particular, 
they prove that: Signed-SAT for  $ \vert \mathcal{T} \vert \geq 3$ is  polynomial
(even trivial),
if $  \bigcap_{S \in \mathcal{S}} S \neq \emptyset $ and NP-complete otherwise. 

\vspace{.05cm}
On the other hand,
Signed-2-SAT is polynomial if, and only if, $ \mathcal{S} $ fulfills the Helly property 
(every sub-family $ \mathcal{V} \subset \mathcal{S} $ satisfying 
$  \bigcap_{S \in \mathcal{V}} S = \emptyset $  contains two sets $ S, S' \in \mathcal{V} $
such that $ S \bigcap S' = \emptyset $) and NP-complete
otherwise. For the case when $ \mathcal{S} $ has the Helly property, i.e. the 
polynomial case, Chepoi et al. 
  show that the satisfiability can in fact be checked in linear time
in the spirit of  the result for classical  2-SAT \cite{AspvallPlassTarjan79}. Also,
they prove that  the Helly property itself can be checked in polynomial time.


 \subsection{Signed Non-Clausal} 

 Signed non-clausal formulas are  defined in \cite{MurrayR91} as negation-free Boolean formulas
with signed literals or signed formulas as atoms. Lehmke \cite{Lehmke96} observed that
 every formula of infinite-valued
\L ukasiewicz logic can be expressed in signed NC, provided that \L ukasiewicz  sum and product are used
instead of classical disjunction and conjunction (this process can blow up a formula exponentially). 
Next we discuss the three approaches to solve the satisfiability problem of 
 signed  non-clausal formulas
  published so far.

 \vspace{.1cm}
(1)   Murray \& Rosenthal's dissolution was first available in classical \cite{MurrayRosenthal93}
and then in finite-valued \cite{MurrayRosenthal93A} logics.  The dissolution rule selects
in a signed NC formula an implicitly conjunctively connected pair of literals 
$ \mathcal{S} \cdot P $, $ \mathcal{S}' \cdot P $ and restructures it in such a 
way that at least one conjunct occurrence of $ \mathcal{S} \cdot P $, $ \mathcal{S}' \cdot P $
is replaced with $  \mathcal{S} \cap \mathcal{S'} \cdot  P$. Producing  $ \emptyset \cdot P $
leads to obvious simplifications such that any unsatisfiable formula is reduced to the
empty formula after a finite number of dissolution steps. The authors in \cite{BeckertHE98} proposed a method to eliminate some redundancies in an input  signed NC formula. No  complexity issues of path
dissolution have been studied.

 \vspace{.1cm}
(2)  Non-Clausal Resolution for classical logic, proposed in \cite{Murray82} (see also \cite{BachmairG01}),
was extended to many-valued logic by Z. Stachniak \cite{Stachniak96,Stachniak01}. 
The basic idea is to derive
from formula $ \phi(p) $ and $ \psi (p) $ (where $ p $ is an atom occurring in 
$ \phi  $ and $ \psi   $) a new formula $ \phi(\rho ) \vee \psi (\rho) $ for certain
variable-free formulas and then to perform logic-specific simplifications. 
Their view of Non-Clausal Resolution is different from ours, and it seems that their approach 
presents some drawbacks  preventing the definition of Non-Clausal
Unit-Resolution, which indeed had not been proposed. No complexities issues were discussed.

\vspace{.1cm}
(3) The method TAS \cite{AguileraGuzmanOjeda95} computes a simplified DNF of an NC formula.
 The input formula is unsatisfiable iff the result is the empty formula. The efficiency
 of  the method comes from the fact that before each application  of the distributive laws,
 unitary models of sub-formulas are computed and used for simplification. The generalization
 of the TAS method to signed NC formulas has been reported in \cite{AguileraGOV98,AguileraGuzmanOjedaValverde01}. Although many experimental running-times
 were published, no computational complexities were obtained.

 \section{Future Work}  \label{sec:futurework} 

Future work that is likely to receive our attention is divided into 
four main lines (see ({\bf 1}) to ({\bf 4}) below) and each of them is generalized
{\em to several non-classical logics}\footnote{For extensions to 
classical  logics, the interested reader is referred to \cite{imaz2021horn}.}:  
({\bf 1})   defining  the Horn-NC class $\mathcal{H_{NC}}$ and   
the   Non-Clausal Unit-Resolution  calculus  $\mathit{UR_{NC}}$, 
and  proving the completeness of each  $\mathit{UR_{NC}}$
for its  $\mathcal{H_{NC}}$; 
({\bf 2}) applying $\mathcal{H_{NC}}$ and        $\mathit{UR_{NC}}$ obtained in ({\bf 1}) to  
NC logic programming;  
({\bf 3})  developing NC  DPLL-based approximate  reasoning  using
 $\mathcal{H_{NC}}$ 
and        $\mathit{UR_{NC}}$  obtained in ({\bf 1});
and ({\bf 4}) establishing Non-Clausal Resolution.

%

%

\vspace{.15cm}
\noindent {\bf (1a) Reg-Horn-NC-SAT.}    Since  $\mathcal{RH}$
should  play in NC form  a   r\^ole  similar to that of   Horn    in   clausal form, 
  worthy research efforts  remain  to  devise a highly-efficient Reg-Horn-NC-SAT algorithm. 
  Here we have already shown that  Reg-Horn-NC-SAT is polynomial and our next goal
   will be proving   that  its complexity is close to  
   $ O(n  \log n) $, i.e.  the one of 
   Reg-Horn-SAT \cite{EscaladaManya94,Hahnle96,Hahnle01,Hahnle03}. 

\vspace{.15cm}
\noindent {\bf (1b)  Lattice-Regular Logic.} We think that some of the classes
published (see related work) for regular Horn formulas when the truth-value set
$ \mathcal{T} $ is partially 
ordered as a lattice  can
be lifted to NC. Specifically, we will study different 
modular and distributive  lattices and define classes $\mathcal{H_{NC}}$ of Horn-NC formulas. Then we will
define  Non-Clausal Unit-Resolution calculi and prove their completeness for 
their corresponding $\mathcal{H_{NC}}$. Then we will verify which clausal
classes preserve their polynomiality when lifted to the NC level.

\vspace{.15cm}
\noindent  {\bf (1c) \L ukasiewicz Logic.}  
SAT-checking the   infinite-valued  \L ukasiewicz Horn  class, 
 $\mathcal{L}_{\infty}$-Horn,  was   
 proved  to be {NP}-Complete  \cite{BorgwardtCP14} and  polynomial  \cite{BofillManyaVidalVillaret15b,  BofillManyaVidalVillaret19} for the 3-valued class, $\mathcal{L}_3$-Horn.
 Thus, our first goal is  to  NC lift  
  $\mathcal{L}_3$-Horn  and  
   determine   the class  $\mathcal{L}_3$-$\mathcal{H_{NC}}$. 
 Then we will analyze whether tractability is preserved in  NC, that is,
 whether SAT-checking $\mathcal{L}_3$-$\mathcal{H_{NC}}$ is polynomial. For that purpose, 
 a former step  is to define the calculus $\mathcal{L}_3$-$\mathit{UR_{NC}}$. 
 We will then deal 
  with $\mathcal{L}_{\infty}$ and try to define
   $\mathcal{L}_\infty$-$\mathcal{H_{NC}}$ and $\mathcal{L}_\infty$-$\mathit{UR_{NC}}$. 
 
 \vspace{.15cm}
\noindent  {\bf (1d) Possibilistic Logic.}   Surveys of this logic
and its numerous applications  are in \cite{DuboisP94,DuboisP04a,DuboisP14}. 
 In possibilistic logic, rather than testing  satisfiability, the deductive problem comes to 
determine the inconsistency degree of   possibilistic conjunctive formulas. 
Such problem for the  necessity-valued Horn clausal class, N$\alpha$-Horn, is polynomial 
\cite{Lang00}, and its complexity, indirectly  discussed in  \cite{AlsinetG00}
via possibilistic logic programming,  is  $ O(n \log n) $. So our first goal
  will be defining the   necessity-valued Horn-NC conjunctive class, or N$\alpha$-$\mathcal{H_{NC}}$,
and attempting to prove that 
computing its inconsistency degree is indeed polynomial. If so, it
 would make N$\alpha$-$\mathcal{H_{NC}}$   the first tractable possibilistic class in NC form. 
Ulterior research is planned to deal with NC formulas
when both necessity and possibility measures are available \cite{LangDP91,Hollunder95,Lang00}.
  Possibilistic paraconsistent reasoning is highly developed
 in the clausal setting \cite{DuboisP14,DuboisP15,CayrolDT18}, and thus,
  taking such advances  as reference,
 new issues are open regarding expressiveness and computing aspects of
   possibilistic NC and HNC formulas.  

\vspace{.15cm}
\noindent  {\bf  (2)  Logic Programming.} Once   
$\mathcal{H_{NC}}$ and $\mathit{UR_{NC}}$  have been defined for any of the four above logics, and
the next step consists of devising   Horn-NC-SAT algorithms for such four logics.
 The expected efficiency of  such  
 algorithms      can have a positive impact on  NC logic programming 
  based on some non-classical logics. Indeed, along the lines of Section \ref{sec:logicprogram},  
we will   define a highly-rich  logic programing language,
where bodies and heads  are NCs with slight syntactical restrictions and
 which can be  efficiently interpreted  using the previously designed  Horn-NC-SAT  
 algorithms. 

 \vspace{.15cm}
\noindent  {\bf (3) DPLL-Based Reasoning.} A   Horn-NC-SAT algorithm is
 indeed the procedure called NC Unit-Propagation which is essential
  in the DPLL skeleton,  proved so far to be the most efficient in propositional logic. 
  Hence, we think that an important axis for future research is the constructing
  of NC DPLL-based reasoners to satisfiability solving and 
  theorem proving and for the   non-classical logics in (1a) to (1d) above.

 \vspace{.15cm}
\noindent {\bf (4) Resolution.} N. Murray in the 1980s \cite{Murray82}  
proposed Non-Clausal Resolution for classical logic (see also the handbook \cite{BachmairG01})  
    in a combined  manner, namely   each NC Resolution application must be followed
    by      logical functions to simplify   the inferred formulas.  NC Resolution has also been 
    extended  to some non-classical logics such as  
 multi-valued logic \cite{Stachniak96,Stachniak01}, 
 fuzzy logic \cite{Habiballa12} and fuzzy description logic \cite{Habiballa07}. 
    Thus a  non-functional  definition 
 of  NC Resolution is missing so far.  However,
  our presented  approach has allowed us to establish  
  Regular NC Unit-Resolution   in a non-functional   
    classical-like fashion (and similarly  for propositional logic in \cite{imaz2021horn}). So we think that our
     approach  can be resumed 
   towards defining  NC Resolution  for  some  logics  including the four  mentioned above in (1a) to (1d).


 \section{Conclusion}

Towards characterizing the first polynomial class in  multi-valued logic
and non-clausal (NC)  form, firstly we have defined
the  class of Regular  Horn-NC formulas,  $\mathcal{RH}$, by means of 
both an inductive, compact function
and of a convenient merging of the  regular Horn and NC  classes. 

As second contribution, we have analyzed  the 
 relationships between $\mathcal{RH}$ and the  classes
regular Horn and regular NC, and in this respect, we have proved that: (i)  $\mathcal{RH}$ syntactically subsumes the Horn class 
but  both classes   are  semantically equivalent;
and (ii) $\mathcal{RH}$ includes all regular NC formulas  whose clausal form is Horn.

 Our third outcome includes both the definition of   Regular Non-Clausal Unit-Resolution, 
 or $\mathit{RUR}_{NC}$,
and  the proof   that  $\mathit{RUR}_{NC}$  is complete for   $\mathcal{RH}$
and  tests  $\mathcal{RH}$  satisfiability in polynomial time.  Hence, the latter fact shows that
our initial goal, that is, finding a
polynomial NC class  beyond propositional logic, is accomplished.

\vspace{.05cm}
We have also discussed how   NC  logic programming can benefit from  our results, arguing
that  classical Horn  rules 
can be notably extended by considering HNC rules  in which
bodies and heads are NCs fulfilling some syntactical constraints and that such syntactical enrichment
is accompanied by a polynomial efficiency in query-answering.

\vspace{.05cm}
Finally, we have discussed several future research lines: (i) defining both
 the Horn-NC class  $ \mathcal{H_{NC}} $
and the  Non-Clausal Unit-Resolution  calculus  $\mathit{UR_{NC}}$ for several logics; 
(ii) developing NC logic programming based on $ \mathcal{H_{NC}} $
and  $\mathit{UR_{NC}}$ for several logics;
(iii)  enhancing NC DPLL-based approximate  reasoning via $ \mathcal{H_{NC}} $
and  $\mathit{UR_{NC}}$;  and (iv)  establishing  Non-Clausal Resolution.

 \vspace{.3cm}
\noindent {\bf Funding Source.} Spanish project ISINC (PID2019-111544GB-C21).

 \vspace{.25cm}
\noindent {\bf \Large  A \ \  Proofs} \label{sect:HNCH*NC}

\vspace{.15cm}
\noindent Before  Theorems  \ref{theorem:first} and \ref{prop:sem-NC},  
a preliminary theorem  (below) is required. 
Thus, we supply successively the next proofs:
%
    Preliminary Theorem, 
     Theorem   \ref{theorem:first}, and  
     Theorem \ref{prop:sem-NC}.

\vspace{.35cm}
\noindent {\bf Preliminary Theorem.}   

\vspace{.15cm}
{\em 
\noindent {\bf Theorem.}   Let  $\varphi$ be an  NC disjunction 
$(\vee  \ \varphi_{1}   \ldots \varphi_i \ldots \varphi_k \,)$. 
   $cl(\varphi) \in \mathcal{H}$    iff  $\varphi$ has  $k-1$  negative  disjuncts and one
disjunct s.t. $cl(\varphi_i) \in \mathcal{H}$, 
 formally     
$$cl(\,(\vee  \ \varphi_{1}   \ldots \varphi_i \ldots \varphi_k \,)\,) 
\in \mathcal{H}$$
\hspace{6.cm} $\mathrm{if \ and \ only \ if}$
 $$(1) \  \exists i,   \mbox{\,s.t.}
  \ cl(\varphi_i) \in \mathcal{{H}} \   \mbox{\,and}   
\ \,(2) \ \mbox{for all} \ j \neq i, \varphi_j \in \mathcal{N}^-.$$    
}

\vspace{-.3cm}
\begin{niceproof} {\em If-then.} By refutation: let     
 $cl(\,(\vee  \ \varphi_{1}   \ldots  \varphi_i \ldots  \varphi_k)\,) \in  \mathcal{H}$ and prove that if   (1) or (2)  are  violated, then  $cl(\varphi)  \notin \mathcal{H}$. 

\begin{enumerate}

\item   [$\bullet$]  Statement (1).  

\begin{enumerate}

\item [$-$] If we take  the case $k=1$, then $\varphi = \varphi_1$.
\vspace{.15cm}
\item  [$-$]  But  $cl(\varphi_1) \notin \mathcal{H}$  implies 
  $cl(\varphi) \notin  \mathcal{H}$.
\end{enumerate} 

\item  [$\bullet$] Statement (2). 

\begin{enumerate}
\item [$-$]  Suppose that, besides $\varphi_i$,   one $\varphi_j,   \,j \neq i$,   has positive literals too.

\vspace{.15cm}
\item  [$-$]  We take a simple case, concretely $k=2, \,\varphi_1=A$
and  $\varphi_2=B$.

\vspace{.15cm}
\item  [$-$] So, $\varphi=(\vee \ \varphi_1 \ \varphi_2) = (\vee \  A  \ B)$, which
implies     $cl(\varphi)  \notin \mathcal{H}$.
\end{enumerate}
\end{enumerate}

{\em Only-If.} For simplicity and without loss of generality, we assume  that  

$(\vee \ \varphi_1 \ldots \varphi_{i} \ldots \varphi_{k-1})=\varphi^- \in  
\mathcal{N}^- 
   \mbox{ and } \varphi_k \in \mathcal{RH}$,  
and  prove that 
$$cl(\varphi)=cl(\,(\vee \ \  \varphi_1 \ldots \varphi_i \ldots \varphi_{k-1}  \,\varphi_k) \,)
=cl(\,(\vee \ \varphi^- \ \varphi_k)\,) \in  \mathcal{H}.$$ 



\vspace{.0cm} $-$ To obtain $ cl(\varphi) $, one must obtain first $cl(\varphi^-)$
and $cl(\varphi_k)$, and so

 \vspace{.2cm}
\hspace{1.cm}
 $(i) \ \ cl(\varphi) = cl(\,(\vee \ \  \varphi^-  \  \varphi_k) \,)= cl(\,(\vee \ \ cl(\varphi^-)  \ \ cl(\varphi_k)\,)\,).$

\vspace{.3cm}
 $-$  By definition of $\varphi^- \in \mathcal{N}^-$, 
 
 \vspace{.2cm}
\hspace{0.8cm} $(ii) \ \  cl(\varphi^-)=\{\wedge \ D^-_1   \ldots D^-_{m-1} \,D^-_m\}$; \          
the $D^-_i$'s are negative clauses.

\vspace{.3cm}
  $-$  By definition of $\varphi_k \in \mathcal{RH}$, 

\vspace{.2cm}
\hspace{.8cm} $(iii) \ \ cl(\varphi_k)=\mathrm{H}=\{\wedge \ h_1 \  \ldots h_{n-1} \,h_n \}$; \
  the $h_i$'s are Horn clauses.

\vspace{.3cm}
 $-$ By   ($i$) to ($iii$), \ 
  
\vspace{.2cm} 
\hspace{1.cm} $cl(\varphi) = cl(\,(\vee \ \  \{\wedge \ D^-_1 \   \, \ldots \, D^-_{m-1} \,D^-_m\} \ \ \{\wedge \ h_1 \   \ldots h_{n-1} \,h_n \, \}\,)\,).$

\vspace{.3cm}
 $-$ Applying $\vee\//\wedge$ distributivity to $cl(\varphi)$ and noting $C_{i}=(\vee \ D^-_1 \ h_{i} \,)$, 

\vspace{.2cm}
\hspace{1.cm} $cl(\varphi) = 
 cl(\,\{\wedge \ \ \{\wedge \ C_1   \ldots C_i \ldots C_n\} \ \  
    (\vee \  \, \{\wedge \ D^-_2 \ldots D^-_{m-1}  \,D^-_m \, \} \  \ \mathrm{H} \,)  \,\} \ ).$

\vspace{.3cm}
 $-$ Since the  $C_i=(\vee \ D^-_1 \ h_{i} \,)$'s are  Horn clauses,  

\vspace{.2cm} 
\hspace{1.cm} $cl(\varphi) = cl (\,\{\wedge \ \ \mathrm{H}_1 \ \ 
(\vee \ \ \{\wedge \ D^-_2 \ldots D^-_{m-1} \,D^-_m \, \}   \ \ \mathrm{H}\, ) \,\} \ ).$

\vspace{.3cm}
 $-$ For $j < m$ we have,
 
\vspace{.2cm} 
\hspace{1.cm} $cl(\varphi) = cl( \ \{\wedge \ \mathrm{H}_1 \ \ldots  \mathrm{H}_{j-1}  \mathrm{H}_j   \  \ (\vee \ \ \{\wedge \ D^-_{j+1} \ldots D^-_{m-1} \,D^-_m\}  \ \  \mathrm{H} \,) \, \} \ ).$

\vspace{.3cm}
 $-$ For $j = m$, \ 
   $cl(\varphi) = \{\wedge \ \mathrm{H}_1 \  \ldots \mathrm{H}_{m-1} \,\mathrm{H}_m \     
\mathrm{H} \,\}=\mathrm{H}' \in \mathcal{H}.$

\vspace{.3cm}
 $-$  Hence   $cl(\varphi)   \in  \mathcal{H}$.
\end{niceproof}

\vspace{.25cm}
\noindent{\bf Proof of Theorem  \ref{theorem:first}.}


\vspace{.25cm}
\noindent {\bf Theorem  \ref{theorem:first}.}
We have: \ $\forall \varphi \in \mathcal{{RH}}:  cl(\varphi) \in \mathcal{H} $.  

\begin{niceproof} We use Definition \ref{def:syntacticalNC}  
of $\mathcal{{RH}}$.
  The proof 
    is done by  structural induction  on   the number of recursions  $r(\varphi)$ 
   needed to include $\varphi$ in  $\mathcal{RH}$. 
  We define $r(\varphi)$ as:
\[r(\varphi)= \left\{
\begin{array}{l l l}

  0   &   \varphi=\mathrm{H}.\\
  
1+max\,\{r(\varphi_1), \ldots, r(\varphi_{k-1}),\,r(\varphi_k)\}    &  \varphi=\langle \odot \  \varphi_1 \ \ldots \varphi_{k-1} \  \varphi_k \rangle. \\

\end{array} \right. \]

$\bullet$ {\it Base Case:} $r(\varphi)=0.$ 

\vspace{.25cm}
\quad   \     By definition,   $\varphi= \mathrm{H} $, and  so 
trivially $cl(\varphi) \in \mathcal{H}$.

\vspace{.3cm} 
$\bullet$ {\em Induction hypothesis:} 
$$     \forall \varphi,  \ r(\varphi) \leq n, \ \ \varphi \in\mathcal{RH} \mbox{ \ entails \ } 
cl(\varphi) \in \mathcal{H}.$$

$\bullet$ {\em Induction proof: \ $r(\varphi)=n+1$.}  

\vspace{.2cm}
\quad   According to Definition \ref{def:syntacticalNC}, cases (1) and (2) below arise.

\vspace{.35cm} 
\hspace{.35cm}  (1)    $\varphi=
\{\wedge  \ \varphi_1   \ldots \varphi_{i}  \ldots  \varphi_k\}
 \in\mathcal{RH}$, where $k \geq 1$.

  \vspace{.3cm} 
\hspace{1.2cm} $-$  By definition of $r(\varphi)$, 

\vspace{.25cm}
\hspace{1.8cm}  \quad
$r(\varphi)=n+1$ \,entails \   $1 \leq i \leq k, \ r(\varphi_i) \leq n.$

\vspace{.25cm}
\hspace{1.2cm} $-$  By induction  hypothesis,

\vspace{.25cm}
\hspace{1.8cm}  \quad $\varphi_i \in\mathcal{RH} 
\mbox{\, and  \,} r(\varphi_i) \leq n
  \mbox{  \ entail \ }  cl(\varphi_i) \in \mathcal{H}$.

\vspace{.25cm}
\hspace{1.2cm}  $-$ It is obvious that,

\vspace{.25cm}
\hspace{1.8cm}
 \quad  $  cl(\varphi) = \{\wedge  \ \  cl(\varphi_1) \  
 \ldots  \ cl(\varphi_{i})   \ldots  cl(\varphi_k) \,\}.$
    
 \vspace{.25cm}
\hspace{1.2cm} $-$  Therefore,

\vspace{.25cm}
\hspace{1.8cm} \quad  
$cl(\varphi) =  \{\wedge  \ \mathrm{H}_1 
\ldots \mathrm{H}_{i}  \ldots \mathrm{H}_k\}= \mathrm{H} \in \mathcal{H}$.

\vspace{.35cm}
\hspace{.35cm}    (2)
 $\varphi=(\vee  \ \varphi_1  \ldots \varphi_{i} \ldots \varphi_{k-1} \ \varphi_k) 
 \in\mathcal{RH}$,
where $k \geq 1$.

 \vspace{.25cm}
\hspace{1.2cm}  $-$ By  Theorem \ref{def:syntacticalNC},  line (3),
 
 \vspace{.2cm}
\hspace{1.8cm} \quad     $0 \leq i \leq k-1, \ \varphi_i   \in \mathcal{N}^-$ and   \,$\varphi_k \in\mathcal{RH}.$

 \vspace{.35cm} 
\hspace{1.2cm} $-$ By  definition of $r(\varphi)$,
 
\vspace{.2cm}
\hspace{1.8cm}  \quad   
$r(\varphi) = n+1$ \ entails   \,$r(\varphi_k) \leq n.$

 \vspace{.25cm} 
\hspace{1.2cm} $-$  By  induction  hypothesis,

\vspace{.25cm}
\hspace{1.8cm}  \quad $d(\varphi_k)  \,\leq n$ \,and \,$\varphi_k \,\in\mathcal{RH}$
 \,entail  $cl(\varphi_k) \,\in \mathcal{H}.$

\vspace{.25cm}
\hspace{1.2cm} $-$  By the first theorem,   {\em only-if,} in the Appendix,\  

 \vspace{.25cm}
\hspace{1.8cm} \quad     $0 \leq i \leq k-1, \ \varphi_i   \in \mathcal{N}^-$ and   \,$cl(\varphi_k) \in \mathcal{H}$ entail:

\vspace{.25cm}
\hspace{2.8cm}
 $cl(\,(\vee  \ \varphi_1  \ldots \varphi_{i} \ldots \varphi_{k-1} \ \varphi_k)\, )  \in \mathcal{H}$.

\end{niceproof}


\noindent{\bf Proof of Theorem \ref{prop:sem-NC}.}  
 
 \vspace{.25cm}
\noindent {\bf Theorem  \ref{prop:sem-NC}. }  $\forall \varphi \in \mathcal{N_C}$:    
if  $cl(\varphi) \in \mathcal{H}$  then    $\varphi \in \mathcal{RH}$.

\begin{niceproof} We use Definition \ref{def:syntacticalNC}  
of $\mathcal{{RH}}$.  The next claims  are trivial:   
\begin{enumerate}

\item [--] By Definition \ref{def:NCformulas} of  $ \mathcal{N_C} $,  
$\mathcal{C} \subset \mathcal{N_C}$. 

\item [--] By  Definition \ref{def:syntacticalNC} of $\mathcal{RH}$,   $\mathcal{H} \subset\mathcal{RH}$.

 \end{enumerate}
Now, we define the depth $d(\varphi)$  of $\varphi$    as     
\[d(\varphi)= \left\{
\begin{array}{l l l}

  0     &   \varphi \in \mathcal{C}.\\
1+max\,\{d(\varphi_1), \ldots, d(\varphi_{i}),\ldots \,,d(\varphi_k)\}   &  
\varphi=\langle \odot \  \varphi_1  \ldots \varphi_{i} \ldots  \varphi_k \rangle.\\

\end{array} \right. \]

The proof is   by  structural induction on   $d(\varphi)$.

\vspace{.2cm}
$\bullet$ {\it Base case:} $d(\varphi)=0$ and    $cl(\varphi) \in \mathcal{H}$.

\vspace{.2cm}
\ \  \  $-$  $d(\varphi)=0$ entails  $\varphi \in \mathbb{C}$.

\vspace{.2cm}
\ \  \    $-$ If $\varphi \notin \mathcal{H}$, then $cl(\varphi)  \notin  \mathcal{H}$,  
 contradicting  the  assumption.

\vspace{.2cm}
\ \  \  $-$ Hence  $\varphi \in \mathcal{H}$ and so by Definition \ref{def:syntacticalNC},
 $\varphi \in\mathcal{RH}$.


\vspace{.2cm}
$\bullet$ {\em Inductive hypothesis:}  
$$ \forall \varphi \in \mathcal{N_C}, \ d(\varphi) \leq n \mbox{\ and } 
cl(\varphi) \in \mathcal{H} \mbox{ \ entail \ } \varphi \in\mathcal{RH}.$$

$\bullet$ {\em Induction proof: $d(\varphi)=n+1$.} 

\vspace{.25cm}
\quad   By Definition \ref{def:NCformulas} of $\mathcal{N_C}$, cases  $(i)$ and $(ii)$  below arise.

\vspace{.3cm}
\hspace{.35cm}    $(i)$  \   $cl(\varphi)= 
  cl(\, \{\wedge \  \varphi_{1}   \ldots \varphi_{i} \ldots   \,\varphi_k\} \, )
\in \mathcal{H}$ and $k \geq 1$.

\vspace{.25cm} 
\hspace{1.2cm} $-$ Since $\varphi$ is a conjunction, $1 \leq i \leq k, \,cl(\varphi_i)  \in \mathcal{H}.$
 

 \vspace{.25cm} 
\hspace{1.2cm} $-$  By  definition of $d(\varphi)$,
 
\vspace{.2cm}
\hspace{1.8cm}  \quad   
$d(\varphi) = n+1$ \ entails   \,$1 \leq i \leq k, \,d(\varphi_i) \leq n.$

\vspace{.25cm}
\hspace{1.2cm}  $-$ By induction hypothesis,

\vspace{.2cm}
\hspace{1.8cm} \quad 
$1 \leq i \leq k, \ \,d(\varphi_i) \leq n,  \ cl(\varphi_i) \in \mathcal{H}$
$  \mbox{\ entail\ } \varphi_i \in\mathcal{RH}.$

\vspace{.25cm}
\hspace{1.2cm} $-$  By Definition  \ref{def:syntacticalNC}, line (2),  

\vspace{.2cm} 
  \hspace{1.8cm} \quad $1 \leq i \leq k, \ \varphi_i   \in\mathcal{RH}$ 
  entails $\varphi \in\mathcal{RH}$.

\vspace{.25cm}
 \hspace{.35cm}     $(ii)$   \ $cl(\varphi) = 
 cl(\, (\vee  \ \varphi_{1}  \ldots \varphi_i  \ldots   \varphi_{k-1} \ \varphi_k) \,) \in \mathcal{H}$ 
  and $k \geq 1$.

 \vspace{.25cm} 
\hspace{1.2cm} $-$  By the first  theorem, {\em if-then,}  in the Appendix,

\vspace{.2cm}
\hspace{1.8cm}   \quad    
 $0 \leq i \leq k-1, \ \varphi_i  \in \mathcal{N}^-$ and   \,$cl(\varphi_k) \in \mathcal{H}.$
 
    \vspace{.25cm} 
\hspace{1.2cm} $-$  By  definition of $d(\varphi)$,
 
\vspace{.2cm}
\hspace{1.8cm}  \quad   
$d(\varphi) = n+1$ \ entails   \,$d(\varphi_k) \leq n.$

 \vspace{.25cm} 
\hspace{1.2cm}  $-$ By       induction  hypothesis,

\vspace{.2cm}
\hspace{1.8cm}  \quad   $d(\varphi_k)  \,\leq n$ and $cl(\varphi_k) \in \mathcal{H}$
\,entail   $\varphi_k \,\in\mathcal{RH}.$

\vspace{.25cm}
\hspace{1.2cm}   $-$  By Definition \ref{def:syntacticalNC},  line (3),

\vspace{.2cm} 
  \hspace{1.8cm} \quad  $0 \leq i \leq k-1, \ \varphi_i   \in \mathcal{N}^-$ and  
   \,$\varphi_k \in\mathcal{RH}$ 
   entail:
  
 \vspace{.25cm} 
  \hspace{2.8cm} \quad $(\vee  \ \varphi_{1}  \ldots \varphi_i  \ldots   \varphi_{k-1} \ \varphi_k) = 
  \varphi \in\mathcal{RH}$. 
  
\end{niceproof}

\end{document}